\documentclass{emulateapj}
\usepackage{placeins}
\usepackage{graphicx}
\usepackage{amsmath}
\shorttitle{Particles in Turbulent Disks} \shortauthors{Zhu et al.}

\newcommand{\del}{{\bf \nabla}}
\newcommand{\bmath}[1]{\mbox{\boldmath{$#1$}}}

\newcommand\msunyr{\rm M_{\odot}\,yr^{-1}}
\newcommand\be{\begin{equation}}
\newcommand\en{\end{equation}}

\newcommand\etal{{\rm et al}.\ }

\begin{document}

\title{DUST TRANSPORT IN MRI TURBULENT DISKS: IDEAL AND NON-IDEAL MHD WITH AMBIPOLAR DIFFUSION}

\author{Zhaohuan Zhu\altaffilmark{1, 2},  James M. Stone\altaffilmark{1}, AND Xue-Ning Bai\altaffilmark{2, 3} }

\altaffiltext{1}{Department of Astrophysical Sciences, 4 Ivy Lane, Peyton Hall,
Princeton University, Princeton, NJ 08544, USA}
\altaffiltext{2}{Hubble Fellow.}
\altaffiltext{3}{Center for Astrophysics,
60 Garden St., Cambridge, MA 02138, USA}
\email{zhzhu@astro.princeton.edu }

\begin{abstract}
We study dust transport in turbulent protoplanetary disks using
three-dimensional global unstratified magnetohydrodynamic (MHD) simulations 
including Lagrangian dust particles. 
The turbulence is driven by the magnetorotational instability (MRI) 
with either ideal or non-ideal MHD that includes ambipolar
diffusion (AD). In ideal MHD simulations, the surface density evolution (except for dust that drifts fastest),  turbulent diffusion,
and vertical scale height of dust  can all be reproduced by simple one-dimensoinal and/or analytical models. 
However,  in AD dominated simulations which simulate protoplanetary disks beyond 10s of AU, the  
vertical scale height of dust  
 is larger than previously predicted. To understand this anomaly in more detail, we carry out 
both unstratified and stratified local
shearing box simulations with Lagrangian particles, and find that turbulence in AD dominated
disks has very different properties  (e.g.,
 temporal autocorrelation functions and power spectra) than turbulence in ideal MHD disks, which
leads to quite different particle diffusion efficiency.
For example, MRI turbulence with AD has
a longer correlation time for the vertical velocity, which
causes significant vertical particle diffusion and large dust scale height. 
In ideal MHD the Schmidt numbers ($Sc$) for radial and vertical turbulent diffusion are $Sc_{r}\sim 1$ and $Sc_{z}\gtrsim 3$,
 but in the AD dominated regime both $Sc_{r}$ and $Sc_{z}$ are $\lesssim 1$. Particle concentration in pressure bumps
 induced by MRI turbulence has also been studied.
 Since non-ideal MHD effects
 dominate most regions in protoplanetary disks, our study
  suggests that modeling dust transport in turbulence driven by MRI with non-ideal MHD effects
 is important for understanding dust transport in realistic protoplanetary disks. 

\end{abstract}

\keywords{accretion, accretion disks - astroparticle physics - diffusion - dynamo - magnetohydrodynamics (MHD) - 
instabilities - turbulence - protoplanetary disks - meteorites, meteors, meteoroids -
stars: pre-main sequence - stars: protostars}

\section{Introduction}
Protoplanetary disks are probably turbulent (Hughes et al. 2011). Turbulence leads to mass accretion and builds
the central star within the disk's lifetime (Hartmann \etal 1998). 

{ Understanding solid transport in turbulent
disks is important for interpreting the meteoritic record of our solar system} (see
 Cuzzi \& Weidenschiling 2006 for a review). Significant radial transport
and mixing of solids may be necessary to explain
the presence of Calcium-Aluminum-rich Inclusions (CAI) in chondritic meteorites (e.g., 
Cuzzi \etal 2003; Ciesla 2010), and the diversity of chondrites (e.g., Anders 1964; Zanda \etal 2006; Jacquet \etal 2012). 
Radial transport of solids has also been incorporated in various
protoplanetary disk time-dependent models to understand 
the redistribution of solids in our solar nebulae
and the implications for comets and meteorites
(Cassen 1996, 2001; Gail 2001, 2002; Wehrstedt 
\& Gail 2002; 
Bockelee-Morvan \etal 2002; Hughes \& Armitage 2010; Jacquet 
\& Robert 2013).

{ Solid transport in turbulent disks is also crucial for planet and planetesimal formation (see Johansen \etal 2014 for a review).}
In the direction perpendicular to the disk midplane, dust settling towards the midplane is balanced by turbulent diffusion, which 
determines the thickness of the dust disk 
 (Weidenschilling 1980; Cuzzi \etal 1993;
Carballido \etal 2006). When the dust disk is thin enough, it can lead to the gravitational
collapse of solids into planetesimals (Safronov 1969; Goldreich \& Ward 1973; 
Youdin \& Shu 2002; Ward 2000; Youdin 2005a, 2005b). 

{ Dust transport also has important implications for protoplanetary disk observations.}
Dust vertical settling in disks helps to explain the 
 spectral energy distributions of protoplanetary disks
(D'Alessio \etal 2006; Furlan \etal 2006; Pinte \etal 2008). 
The radial drift of dust relative to the gas could explain
the dramatically different disk structures revealed by
recent near-infrared (near-IR) polarization imaging and 
submillimeter (submm) inteferometry (Dong \etal 2012; Zhu \etal 2012; Follette \etal 2013;
Andrews \etal 2012; Qi \etal 2013). 

{ Theoretical models on dust transport in turbulence have been developed over the years (Voelk \etal 1980; Markiewicz \etal 1991).}
The turbulent diffusion coefficient for dust particles was derived to understand the thickness of 
dust layers 
(Cuzzi \etal 1993; Dubrulle \etal 1995; Schrapler \& Henning 2004; Carballido \etal 2006). 
To study
particle collision, coagulation, and fragmentation, a 
model to derive relative velocities between
colliding particles has also been developed 
(Cuzzi \& Hogan 2003; Ormel \& Cuzzi 2007; Carballido \etal 2008). 
A more refined model on particle 
turbulent diffusion including 
 particle orbital dynamics in Keplerian disks
(e.g., epicycles and vertical oscillations) has been 
provided by Youdin \& Lithwick (2007).  

{ Numerical simulations have also been carried out to study
dust diffusion in disks with turbulence driven by the magnetorotational instability (MRI, Balbus 
\& Hawley 1991).}
Using unstratified MHD shearing box simulations, 
Johansen \& Klahr (2005) find that the vertical diffusion coefficient is lower than
the turbulent viscosity (which is defined as the total $r-\phi$ turbulent stress normalized by
the local density and orbital frequency), while the radial diffusion coefficient is slightly larger than
the turbulent viscosity. On the other hand, when strong net vertical magnetic fields are imposed, 
both the radial and vertical diffusion coefficients can be significantly 
smaller than the turbulent viscosity (Carballido \etal 2005; Johansen \etal 2006).
The vertical settling of small and large
particles in turbulent disks have been studied using both 
analytical methods and numerical simulations (Fromang \& Papaloizou 2006; Carballido \etal 2006; Turner \etal 2010). 
The radial diffusion coefficients of dust particles in MRI turbulence have 
also been directly measured in simulations (Carballido \etal 2011), which
confirm the analytical formulae suggested by Youdin \& Lithwick (2007).
Global simulations have also been carried out to study dust transport
in global disks.
Fromang \& Nelson (2009) suggest 
that the diffusion coefficients are higher at larger
$z$ since the velocity fluctuations increase significantly at the disk upper layers. 
Particles can also be trapped in the large-scale disk structures induced by 
MHD turbulence (Lyra \etal 2008), such
as vortices (Fromang \& Nelson 2005) 
and zonal flows (Johansen \etal 2009).

{ However, almost all these previous numerical simulations assume ideal MHD.}
In a realistic protoplanetary disk, ideal MHD is a good approximation only
within 0.1 AU where the ionization fraction is large.
Non-ideal MHD effects (e.g., Ohmic dissipation, ambipolar diffusion, and the Hall effect) play essential roles
in disks beyond 0.1 AU (see Armitage 2011; Turner \etal 2014 for a review). From
 0.1 AU to several AU,
a MRI ``dead zone'' due to Ohmic resistivity may exist (Gammie 1996). Dust settling in the ``dead zone'' has been
explored by Fromang \& Papaloizou (2006),
Turner \etal (2010), and Okuzumi 
\& Hirose (2011). 
 In the outer disks beyond 10s of AU, ambipolar diffusion (AD) dominates 
(Bai 2011a, 2011b; Perez-Becker \& Chiang 2011a, 2011b).  Although the effects of AD on MRI turbulence have
been studied both analytically (Blaes \& Balbus 1994; Kunz \& Balbus 2004) and through numerical 
simulations (Mac Low \etal 1995; Brandenburg et al. 1995; Hawley \& Stone 1998),
 dust stransport
in such regions has not been explored yet.  { Since current millimeter observations (e.g. $CARMA$, $SMA$, $ALMA$, $EVLA$) are sensitive
to dust in disks at 10s of AU, studying dust transport in turbulent disks dominated by AD is very important. }

{ In this paper, we study dust transport in MRI turbulent disks in the AD dominated regime for the first time.} 
We will show that the properties of turbulence induced by the MRI with AD can be dramatically different from turbulence induced by 
the MRI with ideal MHD, which has large effects on particle transport in protoplanetary disks.

{ By carrying out global simulations with Lagrangian particles, we also 
test if a simple one-dimensional (1-D) time-dependent model for the dust disk
can reproduce the evolution of dust in three-dimensional (3-D) simulations.} 
If simple 1-D models can be justified, the evolution of dust can be studied in long timescales
without the need of expensive 3-D MHD simulations (e.g. simple models in Birnstiel et 
al. 2013). 
These global MHD simulations also enable us to study particle trapping in the zonal flows.

In Section 2, the analytical theory of particle transport 
in turbulent disks is reviewed. In Section 3, we describe our numerical setup. 
Our results are presented in Section 4. In Section 5, shearing box simulations are carried out to understand the differences
between particle diffusion in ideal and non-ideal MHD with AD. 
A short discussion is given in Section 5, and conclusions are
drawn in Section 6. 

\section{Theory on Particle Transport in Turbulence}
When the gas disk evolves, dust particles will not only follow the gas but also drift, diffuse, and settle in the gas disk.
In the radial direction, the surface density of the dust follows the contaminant equation (Morfill \& Voelk 1984)
\begin{equation}
\frac{\partial \Sigma_{d}}{\partial t}+\frac{1}{r}\frac{\partial}{\partial r}[r(F_{diff}+\Sigma_{d}v_{d,r})]=0\,,\label{eq:sigd}
\end{equation}
where $F_{diff}$ is the radial mass flux of dust due to turbulent diffusion, and $\Sigma_{d}v_{d,r}$
is the mass flux due to dust radial drift from gas-drag. $F_{diff}$ can be written as
\begin{equation}
F_{diff}=-D_{d,r}\Sigma_{g}\frac{\partial}{\partial r}\left(\frac{\Sigma_{d}}{\Sigma_{g}}\right)\,,\label{eq:diff}
\end{equation}
where $D_{d,r}$ is the dust diffusion coefficient in the radial direction, and $\Sigma_{d}$ and $\Sigma_{g}$ are 
the dust and gas surface density. The dust radial velocity
$v_{d,r}$ due to gas-drag is
\begin{equation}
v_{d,r}=\frac{v_{g,r}T_{s}^{-1} - \eta v_{K}}{T_{s}+T_{s}^{-1}}\,,\label{eq:driftv}
\end{equation} 
where $v_{K}$ is the Keplerian velocity, and $\eta$ is the ratio between the pressure gradient and the gravitational force. In an unstratified disk, we have
$\eta$ = $-(r\Omega^{2}\Sigma_{g})^{-1}\partial P/\partial r$. $T_{s}$ is the dimensionless form
of the dust stopping time $t_{s}$ ($T_{s}\equiv t_{s}\Omega$).
We can also incorporate the diffusion term of Equation (\ref{eq:sigd}) into the dust velocity so that the equivalent total dust velocity is
\begin{equation}
v_{d,tot}=\frac{v_{g,r}T_{s}^{-1} - \eta v_{K}}{T_{s}+T_{s}^{-1}}-\frac{D_{d,r}}{r}\frac{d \rm{ln}(\Sigma_{d}/\Sigma_{g})}{d \rm{ln} r}\,\label{eq:driftv3}\,,
\end{equation}
where the last term is the velocity due to turbulent diffusion.

In the vertical direction, dust particles also settle in the disk following the equation of motion
\begin{equation}
\frac{d v_{d,z}}{dt}=-\Omega^{2}z_{d}+\frac{v_{g,z}-v_{d,z}}{t_{s}}\,,\label{eq:vzt}
\end{equation}
where the first term one the right is the vertical gravity toward the disk midplane, and the second term on the right
is the acceleration due to gas drag. In this equation,
$v_{g,z}$ and $v_{d,z}$ are vertical velocities of the gas and dust particles at the particles' vertical positions $z_{d}$. 
When the dust stopping time $t_{s}\ll \Omega^{-1}$ and the gas disk is stationary ($v_{g,z}=0$), 
the particle reaches the terminal velocity of $v_{d,t}=-\Omega^{2}t_{s}z_{d}$
before the particle falls to the disk midplane by gravity. 

If the disk is turbulent ($v_{g,z}\neq$0),
particles can be lifted off the disk midplane by turbulent diffusion, and 
the vertical structure of the dust disk can reach a steady 
state when
the mass flux due to vertical settling balances the mass flux due to turbulent diffusion,
\begin{equation}
\rho_{d}v_{d,t}=D_{d,z}\rho_{g}\frac{\partial}{\partial z}\frac{\rho_{d}}{\rho_{g}}\,,\label{eq:vzt2}
\end{equation}
where $\rho_{g}$ and $\rho_{d}$ are the gas and dust density along the $z$ direction, and
$D_{d,z}$ is the dust diffusion coefficient in the vertical direction.
When the gas disk has a Gaussian density profile as $\rho_{g}(z)=\rho_{{\rm mid},g}{\rm exp}(-z^{2}/2 h^2)$ where $h$
is the scale height of the gas disk $h\equiv c_{s}/\Omega$, Equation (\ref{eq:vzt2}) can be solved to derive
the vertical density profile of the dust as $\rho_{d}(z)=\rho_{{\rm mid},d}{\rm exp}(-z^{2}/2 h_{d}^2)$
with a scale height of
\begin{equation}
h_{d}=\frac{h}{\sqrt{h^{2}\Omega^{2}t_{s}/D_{d,z}+1}}\,.\label{eq:hds}
\end{equation}
If the gas is unstratified (e.g., $\rho_{g}$ is a constant along the $z$ direction), 
as in our unstratified global and local simulations, we can solve Equation (\ref{eq:vzt2}) and
the dust  in disks has a scale height of
\begin{equation}
h_{d}=\sqrt{\frac{D_{d,z}}{\Omega^{2}t_{s}}}\,.\label{eq:hdus}
\end{equation}

At this point, the evolution of dust is fully determined by the
radial and vertical turbulent diffusion coefficients---$D_{d,r}$ and $D_{d,z}$---from Equations (\ref{eq:sigd}), (\ref{eq:hds}) and (\ref{eq:hdus}). 
These turbulent coefficients play essential roles in the
evolution of dust, similar to the important role of turbulent viscosity in the gas disk evolution.

The turbulent diffusion coefficients have been derived in
various works (Volk \etal 1980; Markiewicz \etal 1991; Cuzzi \etal 1993; Dubrulle \etal 1995; Schrapler \& Henning 2004; Carballido \etal 2006; Youdin \& Lithwick 2007). Here, we adopt the formulation from 
Youdin \& Lithwick (2007) who present
a thorough theoretical model including the orbital dynamics of particles in Keplerian disks. 
To enable comparison with our simulations, we highlight some aspects of their theoretical model below. 

First, we focus on the dust vertical diffusion coefficient $D_{d,z}$ which only involves particle motion in the $z$ direction, and it is defined as
\begin{equation}
D_{d,z}\equiv\frac{1}{2}\frac{d\langle z_{d}^{2}\rangle}{dt}\,.\label{eq:ddz}
\end{equation}
The ensemble average can be taken by using trajectories of either many particles over a short period of time, or one particle over a long time.
When dust is very small  and couples with the gas almost perfectly ($t_{s}\ll\Omega^{-1}$), $D_{d,z}$ also equals the gas diffusion coefficient $D_{g,z}$,
\begin{equation}
D_{g,z}=\frac{1}{2}\frac{d\langle z_{g}^{2}\rangle}{dt}=\int_{0}^{\infty}\langle v_{g,z}(\tau)v_{g,z}(0)\rangle d\tau=\int_{0}^{\infty}R_{zz}(\tau) d\tau \,,\label{eq:dgzf}
\end{equation}
where $v_{g,z}(\tau)$ is the vertical velocity of a gas element (or a tracer particle) at time $\tau$, and 
$R_{zz}(\tau)\equiv\langle v_{g,z}(\tau)v_{g,z}(0)\rangle$ is the auto correlation function for $v_{g,z}$.
In a  steady turbulence, we have $R_{zz}(-\tau)=R_{zz}(\tau)$.
If we define the power spectrum  of turbulence as the Fourier transform for the auto correlation function
\begin{equation}
\hat{E}_{g,z}(\omega)=\frac{1}{2\pi}\int_{-\infty}^{\infty}R_{zz}(\tau)e^{i\omega \tau}d\tau \,,\label{eq:egw}
\end{equation}
we then have
\begin{equation}
R_{zz}(0)=\langle v_{g,z}^{2}\rangle=\int_{-\infty}^{\infty}\hat{E}_{g,z}(\omega)d\omega\,.
\end{equation}
If the ensemble average in $R_{zz}(\tau)$ is defined as averaging $v_{g,z}(t+\tau)v_{g,z}(t)$ of a fluid element over a long time,
i.e., $\lim_{T \to +\infty}1/T\times\int_{-T/2}^{T/2} v_{g,z}(t+\tau)v_{g,z}(t) dt$, we have 
$\hat{E}_{g,z}(\omega)=(2\pi/T) |\hat{v}_{g,z}(\omega)|^2$ by the convolution theorem. 

One important quantity to characterize turbulence is 
 the integral timescale or
the eddy time, defined  as $t_{eddy,z}\equiv\int_{0}^{\infty} R_{zz}(\tau)/R_{zz}(0) d\tau$. If turbulence is isotropic, we have $t_{eddy,z}\sim t_{eddy,x} \sim t_{eddy,y}$ so that a single $t_{eddy}$ is used to denote all three. 
If we set $\omega=0$ in Equation (\ref{eq:egw}) and use Equation (\ref{eq:dgzf}) and the definition of $t_{eddy}$, we have
\begin{equation}
\hat{E}_{g,z}(0)=\frac{1}{2\pi}\int_{-\infty}^{\infty}R_{zz}(\tau)d\tau= \frac{\langle v_{g,z}^{2}\rangle t_{eddy}}{\pi}\, \label{eq:Eg0}
\end{equation}
and thus
\begin{equation}
D_{g,z}=\pi \hat{E}_{g,z}(0) =\langle v_{g,z}^{2}\rangle t_{eddy}\,,\label{eq:dgz}
\end{equation}
which suggests that the diffusion coefficient only depends on the power spectrum at $\omega=0$,
and $D_{g,z}$ is the product of the mean squared velocity and the eddy time.

If we plug $D_{g,z}=\langle v_{g,z}^{2}\rangle t_{eddy}$ into Equations (\ref{eq:hds}) and (\ref{eq:hdus}), the disk scale height for small particles ($t_{s}\ll\Omega^{-1}$) in stratified disks is
\begin{equation}
h_{d}=\frac{h}{\sqrt{h^{2}\Omega^{2}t_{s}/(\langle v_{g,z}^{2}\rangle t_{eddy})+1}}\,,\label{eq:hds3}
\end{equation}
and the dust scale height in unstratified disks is 
\begin{equation}
h_{d}=\sqrt{\frac{\langle v_{g,z}^{2}\rangle t_{eddy}}{\Omega^{2}t_{s}}}\,.\label{eq:eqhd3}
\end{equation}

In order to derive the scale height for particles with any $t_{s}$, Youdin \& Lithwick (2007) solve
the Langevin equation (Equation (\ref{eq:vzt})) using the   
Fourier transform, and  derive
\begin{equation}
\hat{v}_{d,z}=\frac{\omega}{\omega+i t_{s}(\Omega^{2}-\omega^{2})}\hat{v}_{g,z}\,.
\end{equation}
The scale height is 
\begin{eqnarray}
h_{d}^{2}&=&\lim_{T \to +\infty} \frac{1}{T}\int_{-T/2}^{T/2} |z|^2 dt=\frac{2\pi}{ T}\int_{-\infty}^{\infty} |\hat{z}_{d}|^{2} d\omega\nonumber\\
&=&\int_{-\infty}^{\infty} |\hat{z}_{d}|^{2}\frac{\hat{E}_{g}}{|\hat{v}_{g,z}|^2}d\omega\,,\label{eq:eqhd}
\end{eqnarray}
where  $\hat{z}_{d}=i \hat{v}_{d,z}/\omega$.

Up to this point, the detailed form of the power spectrum of turbulence has not been assumed. In order to proceed,
 Youdin \& Lithwick (2007) assume the turbulence power spectrum in a uniform unstratified disk is
\begin{equation}
\hat{E}_{g}(\omega)=\frac{\langle v_{g}^{2}\rangle}{\pi}\frac{t_{eddy}}{1+\omega^{2}t_{eddy}^2}\,,\label{eq:ego}
\end{equation}
so that Equation (\ref{eq:eqhd}) can be integrated to be
\begin{equation}
h_{d}^{2}=\frac{D_{g,z}}{\Omega^2 t_{s}}\frac{1+{\rm St}}{1+{\rm St}+{\rm St} (t_{eddy}\Omega)^2}\,\label{eq:eqhd2}
\end{equation}
where the Stokes number (St) is defined as St$\equiv t_{s}/t_{eddy}$. When $St\ll 1$, Equation (\ref{eq:eqhd2}) 
reduces to Equation (\ref{eq:eqhd3}).
In the opposite limit, when $St\gg1$, 
Equation 
(\ref{eq:eqhd2}) can also be approximated by
Equation (\ref{eq:eqhd3})  (Carballido \etal 2006) 
as long as $t_{eddy}\sim \Omega ^{-1}$. 

Particle diffusion in the radial direction is defined in a similar way as 
that in the vertical direction (Equation (\ref{eq:ddz}))
\begin{equation}
D_{d,x}\equiv\frac{1}{2}\frac{d\langle x_{d}^{2}\rangle}{dt}\,.\label{eq:ddx}
\end{equation}
Then, the rest derivation of $D_{d,x}$ is similar to the derivation above, except that 
it involves epicyclic oscillation, and
both $v_{r}$ and $v_{\phi}$ need to be considered. Assuming the power spectra of $v_{g,x}$, $\delta v_{g,y}$, and $v_{g,z}$ 
have similar forms,\footnote{Youdin \& Lithwick (2007) use the local shearing box approximation so that $x$ is the $r$ direction, and
$y$ is the $\phi$ direction.} Youdin \& Lithwick (2007) derive
the particle radial diffusion coefficient as
\begin{equation}
D_{d,x}=t_{eddy}\frac{\langle v_{g,x}^2\rangle+4 T_{s}^2 \langle \delta v_{g,y}^2\rangle+4T_{s} \langle v_{g,x}\delta v_{g,y}\rangle}{(1+T_{s}^2)^2}\,.\label{eq:ddx1}
\end{equation}
For isotropic turbulence $\langle v_{g,x}^2\rangle=\langle \delta v_{g,y}^2\rangle$, $\langle v_{g,x}\delta v_{g,y}\rangle=0$, Equation 
(\ref{eq:ddx1}) is reduced to
\begin{equation}
D_{d,x}=D_{g,x}\frac{1+4 T_{s}^2}{(1+T_{s}^2)^2}\,.\label{eq:ddx2}
\end{equation}
where $D_{g,x}=\langle v_{g,x}^2\rangle t_{eddy}$.

Finally, if $D_{g,x}$ (Equation (\ref{eq:ddx2})) is given, the evolution of the dust disk surface density (Equation (\ref{eq:sigd}))  is fully determined. 
If $D_{g,z}$ is given, the vertical structure of the dust disk is also determined (Equation (\ref{eq:eqhd2})). To get $D_{g,x}$ and $D_{g,z}$,
we need to know $\langle v_{g,x}^2\rangle$, $\langle v_{g,z}^2\rangle$ and $t_{eddy}$.
Conveniently, previous ideal MHD simulations suggest that $t_{eddy}$ is simply around 
$\Omega^{-1}$ (Fromang \& Papaloizou 2006; Carballido \etal 2011).

In the following, we will study the surface density evolution and vertical structure of the dust disk in our simulations, and compare the results with 
those derived from the simple one-dimensional (1-D) and analytical models (Equations (\ref{eq:sigd}) and (\ref{eq:eqhd3})).
Although the 1-D and analytical models  that assume $t_{eddy}\sim \Omega^{-1}$
can reproduce dust distribution in ideal MHD simulations very well, $t_{eddy}\sim \Omega^{-1}$
breaks down in MHD simulations with AD (\S 4). Then we perform local shearing box simulations  
to directly measure $t_{eddy}$ and the power spectrum of turbulence, which
will  be compared with
the power spectrum normally assumed  (Equation (\ref{eq:ego})) (\S 5).

\section{Simulations}

The gas dynamics is computed using Athena (Stone \etal 2008), a higher-order
Godunov scheme for hydrodynamics and magnetohydrodynamics using the piecewise 
parabolic method for
spatial reconstruction (Colella \&Woodward 1984), the corner transport upwind method 
for multidimensional integration (Colella 1990), and constrained transport
 to conserve the divergence-free property for magnetic fields
(Gardiner \& Stone 2005, 2008).  Cylindrical grids (Skinner \& Ostriker 2010) are used to
simulate MRI turbulence in global disks.

We use the particle 
integrator in Athena (Bai \& Stone 2010; Zhu \etal 2014) to study the motion of dust particles in MRI turbulent disks. 
Dust particles are implemented as Lagrangian
particles with the
dust-gas drag term, following
\begin{equation}
\frac{d\mathbf{v}_{i}}{dt}=\mathbf{f}_{i}-\frac{\mathbf{v}_{i}-\mathbf{v}_{g}}{t_{s}}\,,\label{eq:motion}
\end{equation}
where  $\mathbf{v}_{i}$ and $\mathbf{v}_{g}$ are the velocity vectors for particle $i$
and the gas, $\mathbf{f}_{i}$ is the gravitational force from the central star, 
and $t_{s}$ is the dust stopping time due to gas drag. 
This equation is integrated with time using second-order integrators which can preserve the geometric
properties of particle orbits (Zhu \etal 2014). 
Since in a real disk, the vertical gravity is very small close to the disk midplane, the unstratified disk setup is a good approximation for simulating 
 the gas dynamics there (e.g. within half the gas scale height of the disk). However, even close to the disk midplane, dust can 
 have a significant stratification since it can have a much smaller scale height than the gas
 due to dust settling.
Thus, we  include the vertical gravitational force for dust particles, in which case dust is
allowed to settle toward the disk midplane.

\subsection{The Gas Disk}

\begin{table*}[ht]
\begin{center}
\caption{Global Simulations \label{tab1}}
\begin{tabular}{ccccccccccc}
\tableline
Run & Resolution  & B &  t &Par. No. & $\langle v_{g,r}^{2}\rangle$ & $\langle \delta v_{g,\phi}^{2}\rangle$ & $\langle v_{g,z}^{2}\rangle$ & $\langle \rho v_{g,r} \delta v_{g,\phi} \rangle$/$\langle \rho \rangle$&  $\langle -B_{r}B_{\phi}\rangle$/$\langle \rho \rangle$ & $\alpha$ \tablenotemark{a} \\
Name &   & & 2$\pi/\Omega(r_{0})$ &Millions & ($c_{s}^{2}$) &  ($c_{s}^{2}$) & ($c_{s}^{2}$) & ($c_{s}^{2}$) & ($c_{s}^{2}$)    &  \\
\tableline
V1e4 & $576\times1024\times32$ & Vert. $\beta_{0}=$1e4 & 100 & 1& 0.036 & 0.015 & 0.010 & 7.5e-3 & 0.028  & 0.035  \\
V1e5 & $576\times1024\times32$  & Vert. $\beta_{0}=$1e5 & 100 & 1&0.031 & 0.010 & 6.8e-3 & 5.3e-3 & 0.017  & 0.022  \\
T1e2 & $576\times1024\times32$  & Tor. $\beta_{0}=$100 & 100 & 1&0.031 & 0.015 & 0.010 & 7.5e-3 & 0.024  & 0.032  \\
T1e3 & $576\times1024\times32$  & Tor. $\beta_{0}=$1e3 & 100 & 1&0.027 & 0.012 & 7.5e-3 & 6.2e-3 & 0.018  & 0.025  \\
AD1e3 & $576\times1024\times32$  & Vert. $\beta_{0}=$1e3 & 100 & 1 &5.2e-3 & 1.6e-3 & 1.2e-3 & 8.8e-4 & 1.5e-3  & 2.4e-3  \\
AD2.5e4 & $576\times1024\times32$  & Vert. $\beta_{0}=$2.5e4 & 100 & 1&3.9e-3 & 1.9e-4 & 3.4e-5 & 5.0e-4 & 1.8e-4  & 6.4e-4  \\
\tableline
High reso. (32/h) &&&&&&&&&\\
\tableline
V1e4H & $1152\times2048\times64$  & Vert. $\beta_{0}=$1e4 & 60 & 300&0.032 & 0.021 & 0.013 & 7.7e-3 & 0.030  & 0.039  \\
T1e2H & $1152\times2048\times64$  & Tor. $\beta_{0}=$100 & 100 & 1&0.037 & 0.018 & 0.011 & 0.010 & 0.023  & 0.033  \\
AD1e3H &$1152\times2048\times64$  & Vert. $\beta_{0}=$1e3 & 100 & 1& 7.0e-3 & 2.0e-3 & 1.7e-3 & 9.8e-4 & 1.7e-3  & 2.8e-3  \\
\tableline
\end{tabular}
\tablenotetext{1}{$\alpha\equiv\frac{\langle -B_{r}B_{\phi}\rangle}{\langle\rho\rangle c_{s}^2}+\frac{\langle\rho v_{g,r} \delta v_{g,\phi}\rangle}{\langle\rho\rangle c_{s}^2}$. }
\end{center}
\end{table*}

The gas disk  is unstratified so that the gas density is constant in the $z$ direction.  
The initial radial profile of the gas disk is 
\begin{eqnarray}
\rho_{g}(r,\phi,z)=\rho_{g,0}\left(\frac{r}{r_{0}}\right)^{-1}\label{eq:eqini}\\
T(r,\phi,z)=T_{0}\left(\frac{r}{r_{0}}\right)^{-1/2}\,,
\end{eqnarray}
so that $h/r=c_{s}/v_{\phi}=(h/r)_{r=r_{0}}(r/r_{0})^{1/4}$. 
We choose $\rho_{g, 0}=1$, $r_{0}=1$, and $(h/r)_{r=1}=0.1$. The
disk is locally isothermal at each $r$, which means that 
it has the same temperature in the $\phi-z$ plane at a given $r$. 
The local isothermal assumption is valid when the 
viscous/turbulent heating is much less efficient than the stellar irradiation
so that the disk temperature is only controlled by the
stellar irradiation (Kenyon \& Hartmann1987; Calvet \etal 1991). 
Since we are interested in the outer disk beyond 10s of AU, 
where ambipolar diffusion dominates, the
 local isothermal assumption is a good approximation. {To compare with
the 1-D dust surface density evolution equation, we  use $\rho_{g}$ in the unstratified disk to represent
$\Sigma_{g}$, and choose the slope of $\rho_{g}$ as -1 (Equation \ref{eq:eqini}), which is the slope of the surface density in the
$\alpha$ disk similarity solution. 
However, in a realistic stratified disk, the midplane
density has a steeper gradient along the radius than the surface density (e.g. demonstrated in Appendix B of Zhu \etal 2014).
Thus, particles drift significantly slower in our simulations than in a realistic disk. Such slower drift allows 
us to evolve the dust disk for a longer time but also allow more particles concentrate in pressure bumps (\S 6.2). }

Turbulence is driven by the MRI in both ideal and non-ideal
MHD simulations. Following Bai \& Stone (2011), we include 
the effect of AD by modifying the induction equation as
\begin{equation}
\frac{\partial {\bmath B}}{\partial t} = \del \times \left({\bmath v} \times {\bmath B}- \frac{4\pi}{c} \eta_{A}{\bmath J}_{\perp} \right),
\end{equation}
where ${\bmath B}$ is the magnetic field, 
 ${\bmath J} _{\perp}= (\del \times {\bmath B})_{\perp}$ is the component of the current density that is
  perpendicular to the direction of the magnetic field, and $\eta_{A}$ is
the ambipolar diffusivity $\eta_{A}\equiv v_{A}^{2}/(\gamma \rho_{i})$ in which $v_{A}$ is the Alfven speed and 
 $\gamma \rho_{i}$ is the neutral-ion collision frequency.
The effect of AD in disks can be characterized by the dimensionless 
  parameter Am (Chiang \& Murray-Clay 2007),
\begin{equation}
{\rm Am}\equiv \frac{\gamma \rho_{i}}{\Omega}\,,
\end{equation}
which is the number of collisions for a neutral molecule with ions in the dynamical timescale $1/\Omega$.
{ We adopt Am$=1$ which is the typical value for the protoplanetary disk from 10 to 100 AU (Bai 2011a, 2014). }
Thus, the ambipolar diffusivity $\eta_{A}$ becomes $\eta_{A}=v_{A}^{2}/\Omega$. 

For ideal MHD simulations, the disk is threaded by
either net vertical or net toroidal magnetic fields. For AD simulations, we only study disks threaded by net vertical fields since
net toroidal fields generate little turbulence in disks dominated by AD (Bai \& Stone 2011).
The initial magnetic fields have a constant plasma $\beta=8\pi \rho c_{s}^{2}/B^{2}$ everywhere in the whole disk.
The initial configuration of magnetic fields is shown in Table 1. Ideal MHD runs with net vertical and toroidal 
fields have names 
starting with ``V'' and ``T'' respectively, while AD runs start with ``AD''.
For each magnetic field geometry, we have varied the initial plasma $\beta$ by at least one order
of magnitude.   

Our cylindrical grids span from 0.5 to 4 in the $r$ direction, $0$ to $2\pi$ in the $\phi$ direction, and 
-0.1 to 0.1 in the $z$ direction. The vertical domain of $z=[-0.1,0.1]$ is equivalent to 2 $h$ at $r=1$, 
4.8 $h$ at the inner boundary ($r=0.5$) and 0.7 $h$ at the outer boundary ($r=4$).
The grid is uniformly spaced in all $r$, $\phi$ and $z$ directions.  Our standard simulations
have the resolution of $576\times1024\times32$ in the $r$, $\phi$, and $z$ directions, which is 16 grids per $h$ at $r=1$
 in all three directions. 
The high resolution runs with the resolution of $1152\times2048\times64$ have 32 grids per $h$ at $r=1$.

At the outer radial boundary, the physical quantities in ghost zones 
are set to be fixed at the initial values (as in Zhu \etal 2014).
Such a boundary can absorb waves quite efficiently in numerical codes using Godunov-type schemes.
At the inner boundary, the open boundary condition used in Sorathia \etal (2012) has been applied to allow
mass accretion.
Periodic boundary conditions have been applied in both $\phi$ and $z$ directions.

\subsection{Particle Component}
In the initial condition, we 
distribute dust particles in a way that leads to the same radial profile of surface density as the gas (Equation (\ref{eq:eqini})).\footnote{
For passive particles, we can arbitrarily scale the dust surface density without affecting the dynamics. 
To compare with observations,
we only need to scale the dust density using the realistic dust-to-gas mass ratio.}
In detail, to ensure that the surface density of the dust has a slope of -1, each particle is placed in the disk at 
a constant probability in both $r$ and $\phi$ directions. All particles are initially placed at the disk midplane 
with circular Keplerian speeds.

We evolve seven types/species of particles simultaneously with the gas. 
For each particle type,  there are
$10^{6}$ particles.
We assume that
all these particles are in the Epstein regime, so that
 the dust stopping time (Whipple 1972; Weidenschilling 1977, we use the notation from Takeuchi 
\& Lin 2002) is
\begin{equation}
t_{s}=\frac{ s \rho_{p} }{\rho_{g}v_{T}}\,,\label{eq:ts}
\end{equation}
where $\rho_{g}$ is the gas density, $s$ is the dust particle radius, $\rho_{p}$ is the dust particle density (we 
choose $\rho_{p}$=1 g cm$^{-3}$), $v_{T}$=$\sqrt{8/\pi}c_{s}$, and $c_{s}$ is the gas sound speed. 

To make our results general, we do not specify the length and mass unit in our simulations. 
Each particle type in our simulations has one certain size ($s$) in Equation (\ref{eq:ts}).
With our chosen particle size ($s$), the dust stopping time for each particle type at $r=1$ in the initial condition
is shown in Table 2.
Note that  $t_{s}$ will evolve with time in the simulation since
$\rho_{g}$ and $v_{T}$  are changing with time along the particle's trajectory. 

Given a realistic protoplanetary disk structure, particles in our simulations
can be translated to real particles having some physical sizes  in protoplanetary disks.
Using Equation (\ref{eq:ts}) and $c_{s}=h\Omega$, 
a dust particle with size $s$ at the midplane of a realistic disk ($\rho_{mid}=\Sigma/(\sqrt{2\pi}h)$) has
\begin{equation}
T_{s}=t_{s}\Omega=\frac{\pi s \rho_{p}}{2\Sigma_{g}}=1.57\times10^{-3}\frac{\rho_{p}}{1 {\rm g \, cm^{-3}}}\frac{s}{1\, {\rm mm}}\frac{100\, {\rm g \,cm^{-2}}}{\Sigma_{g}}\,.\label{eq:ts3}
\end{equation}
For example, if we are studying an $\alpha$ disk similarity solution with $\alpha=0.01$, $\dot{M}=10^{-8}\msunyr$,
and $ T = 221 (R/{\rm AU})^{-1/2}$~K, the disk surface density is then $\Sigma_{g}=178 (r/{\rm AU})^{-1}$ g cm$^{-2}$. Using 
Equation (\ref{eq:ts3}),
we can derive that the smallest particle type in our simulations with $T_{s}=0.007041$ at $r=1$ corresponds to 
1.5 mm particles at 5 AU and 0.4 mm particles at 20 AU in such a disk.  With different realistic disk structures, 
Table 2  gives the real particle sizes that our simulated particle types correspond to.
To relate our results  with the latest $ALMA$ observation on Oph IRS 48 (van der Marel et al. 2013, Bruderer \etal 2014 ), 
we give the corresponding size of particles in
Oph IRS 48 in the last column.  
However, with a realistic surface density,
some particle types are in the Stokes rather than the Epstein regime. Those particle types can
only be considered as a numerical experiment to explore the effect of a large stopping time on the particle distribution,
rather than a realistic model of such particles.

\begin{table}[ht]
\begin{center}
\caption{Corresponding particle sizes \label{tab1}}
\begin{tabular}{lccccc}
 
\tableline\tableline
Par  & $t_{s}\Omega$ & Par. Size & Par. Size & Par. Size  &  Par. Size \\
                        &   At $r$=1    &   At 5 AU in  a  &   At 5 AU in &  at  20 AU      &   At  30 AU in            \\
                        & Initially & MMSN Disk\tablenotemark{a}  &  $\alpha$ Disk \tablenotemark{b} & In $\alpha$ Disk & Oph IRS 48 \tablenotemark{c} \\
\tableline
a &   0.007041   &  6.8 mm & 1.5 mm & 0.4 mm & 3 $\mu$m \\
b &   0.07041   & 6.8 cm & 15 mm &  4 mm  &  30 $\mu$m  \\
c &   0.7041 & Stokes &  15 cm & 4 cm  & 0.3 mm \\
d &   7.041  &  Stokes\tablenotemark{d} & Stokes & 40 cm   & 3 mm\\
e  &   70.41 &  Stokes &  Stokes  & 4 m & 3 cm \\
f  &   704.1 &  Stokes &  Stokes  & Stokes & 30 cm  \\
g &   7041 &  Stokes &  Stokes  & Stokes & 3 m  \\
\tableline
\end{tabular}
\tablenotetext{1}{ A MMSN disk has $\Sigma_{g}=1700 (r/{\rm AU})^{-1.5}$ g cm$^{-2}$.}
\tablenotetext{2}{ The $\alpha$ disk has $\Sigma_{g}=178 (r/{\rm AU})^{-1}$ g cm$^{-2}$, which is the surface density of a constant $\alpha=0.01$ accretion disk with
$\dot{M}=10^{-8}\msunyr$,
and $ T = 221 (r/{\rm AU})^{-1/2}$~K }
\tablenotetext{3}{ Oph IRS 48 disk with a 2 M$_{\odot}$ central star,  
$\Sigma_{g}=1.92 (r/{\rm AU})^{-1}$ g cm$^{-2}$ and 
$ T = 542 (r/{\rm AU})^{-1/2}$~K (Bruderer \etal 2014).}
\tablenotetext{4}{ Particles in the Stokes regime, see text.} 
\end{center}
\end{table}

The turbulent disk with both gas and dust is evolved for 100 orbits.  Throughout the paper, an orbit 
is defined as 
the orbital period at $r=1$ (2$\pi/\Omega(r=1)$). 
During this time, most of our particle types have reached
vertical equilibrium since the settling timescale of dust particles ($\Omega^{-1}(T_{s}+T_{s}^{-1})$) is only
22 orbits for the smallest particles in our simulations (Par. a).

\section{Results}

\begin{figure*}[ht!]
\centering
\includegraphics[width=0.8\textwidth]{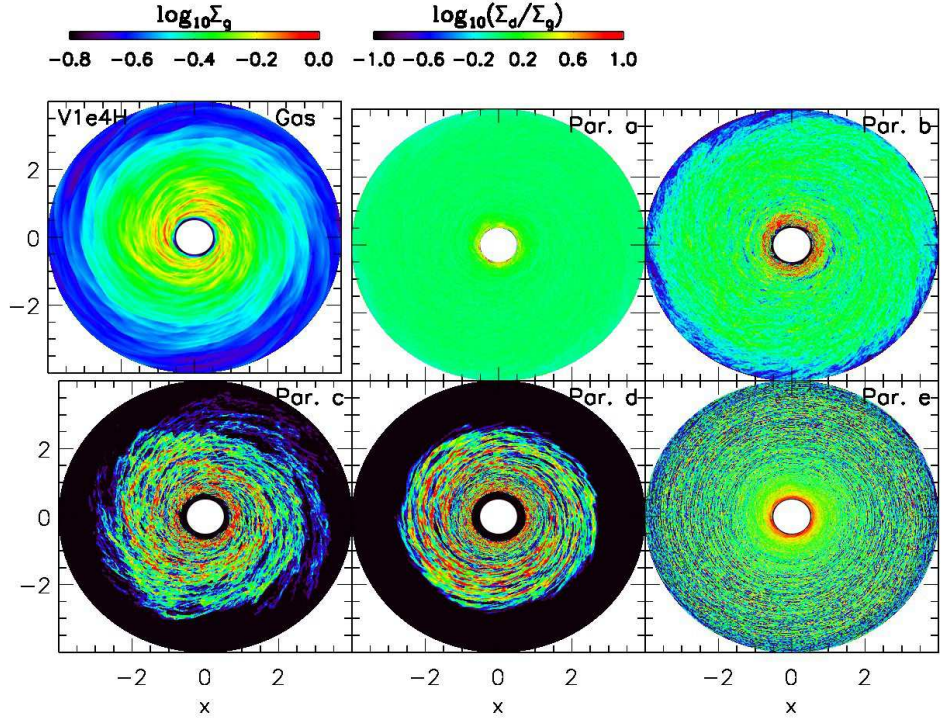} 
\vspace{-0.3 cm}
\caption{Disk surface density at 60 orbits for the ideal MHD simulation V1e4H .
The upper left panel shows the surface density of the gas disk, while
the other five panels show the ratio between the dust and gas surface density.
The smallest particles (Par. a) couple with the gas so well that the density
ratio is close to 1. With
 increasing particle size from Par. a to Par. c, the disk becomes smaller due to the faster radial drift,
 and dust concentrates more within 
the spiral arms.  Par. d and e, with $T_{s}>1$, start to decouple from the gas and the concentration
is more axisymmetric. } \label{fig:gaspar}
\end{figure*}

{ MRI turbulence is significantly suppressed when AD dominates in the disk,
as shown in Table 1.}
In ideal MHD runs,
both V1e4 and T1e2 have similar total stresses which are equivalent to $\alpha\sim0.03$, and V1e5 and T1e3 have
$\alpha\sim0.02$. When AD dominates, even strong vertical
fields ($\beta_{0}=1000$ in AD1e3) can only lead to $\alpha\sim0.002$. This strong suppression
of turbulence by AD is consistent with previous studies (Bai \& Stone 2011).
Since the density, temperature, and magnetic fields
vary radially in our setup, $\alpha$ varies radially too. By measuring
the radial profile of $\alpha$ in our simulations, we find $\alpha\propto r^{-5/4}$.
This trend is consistent with the $\alpha$ predictor found in 
unstratified shearing box simulations (Hawley \etal 1995) that
$\alpha\propto L_{z}\Omega^{2}\lambda_{c}/c_{s}^2$ where $L_{z}$ is
the size of the box in the $z$ direction, and 
$\lambda_{c}\sim 2\pi v_{A}/\Omega\sim \sqrt{\pi/\rho}B/\Omega$. 
With a constant $\beta_{0}$ throughout the disk, we have $B\propto r^{-3/4}$, $\lambda_{c}\propto r^{5/4}$, and 
the predictor above suggests that $\alpha$ is indeed $\propto r^{-5/4}$.

{  Since the mean squared velocities (i.e., $\langle v_{g,r}^{2}\rangle$, $\langle \delta v_{g,\phi}^{2}\rangle$,
$\langle v_{g,z}^{2}\rangle$) together with the eddy time ($t_{eddy}$) determine
the dust diffusion coefficients (e.g. $D_{d,x}$, $D_{d,z}$)
and the dust scale height (Equations (\ref{eq:dgz}), (\ref{eq:eqhd2}), (\ref{eq:ddx1}), (\ref{eq:ddx2})),
we measure the averaged mean squared velocities in all our simulations, as also shown in Table 1.}
All our runs have
$ \langle v_{g,r}^2\rangle \gtrsim \langle \delta v_{g,\phi}^2\rangle \gtrsim \langle v_{g,z}^2\rangle$. 
The ratio between $ \langle v_{g,r}^2\rangle $ and $ \langle v_{g,z}^2\rangle$ is normally 3 in ideal
MHD runs, while this ratio can be significantly larger in AD runs. Another trend  is that 
$\langle v_{g,r}^2\rangle \sim \alpha c_{s}^2$ in ideal MHD runs, while  
$\langle v_{g,r}^2\rangle > \alpha c_{s}^2$ in AD runs.
Nevertheless, since $\langle v_{g,r}^2\rangle$ is several times of $\langle v_{g,z}^2\rangle$ in all our runs, we expect that 
MRI turbulence should lead to larger radial diffusion coefficients than vertical diffusion coefficients.
 However, as shown in \S 4.2 and \S 5, 
this expectation is not true for turbulence generated by the MRI with AD due to the large edgy time for
$v_{z}$ in these disks.

{  During 100 orbits, the dust distribution has evolved significantly due to both radial drift and
turbulent diffusion.}
Figure \ref{fig:gaspar} shows the disk surface density for the gas and different types of particles at 60 orbits in V1e4H. In
particle panels, the ratio of the dust to the gas surface density is shown.
The smallest particles (Par. a) couple with the gas so well that the density ratio
is almost 1 throughout the disk. Bigger particles 
(e.g., Par. b and c) which drift faster in disks
have  smaller radial extents than the gas. 
They are also more concentrated in spiral
arms, suggesting that particles drift azimuthally responding to the gas turbulent  structures. 
Particles with $T_{s}>1$ (e.g.,
Par. d and e) start to decouple from the gas. Due to their long coupling times with the gas, 
they cannot respond to
density fluctuations occurring faster than the orbital timescale. Thus, 
the inhomogeneity of the gas fluctuations in the azimuthal direction (e.g. spiral arms) only 
has an average effect on them,  causing more axisymmetric particle distributions.  
In the following, we will study particle radial drift, and vertical settling in more detail. 

\begin{figure*}[ht!]
\centering
\includegraphics[width=0.76\textwidth]{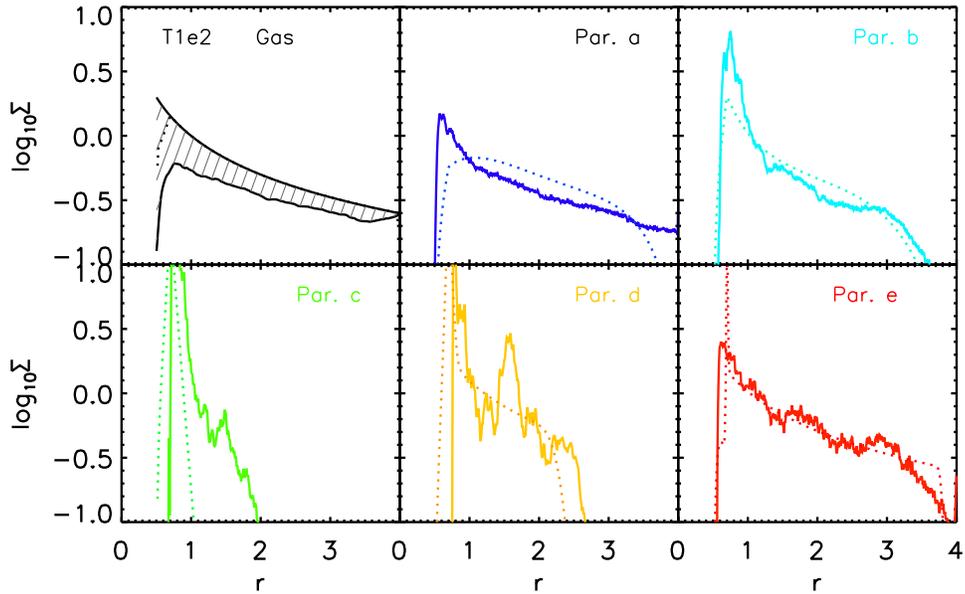} 
\vspace{-0.3 cm}
\caption{Azimuthally
averaged disk surface density for the gas (black curves) 
and different types of particles (colored curves) in
run T1e2 at 100 orbits. In the gas panel,  the upper black solid curve is the initial gas surface density,
while the lower black solid curve is the gas surface density at 100 orbits. The shaded region is thus due to disk accretion. The dotted curves
are from the 1-D dust  model
(Equation (\ref{eq:sigd})) with the fixed $\Sigma_{g}$ (Equation (\ref{eq:sigg0})) and $D_{d,r}$. This 1-D model
generally agrees with simulation results, but it  leads to a too small disk for Par. c, and
 unable to reproduce large-scale density peaks for Par. d and Par. e.
} \label{fig:radialtorpaper1}
\end{figure*}

\begin{figure*}[ht!]
\centering
\includegraphics[width=0.76\textwidth]{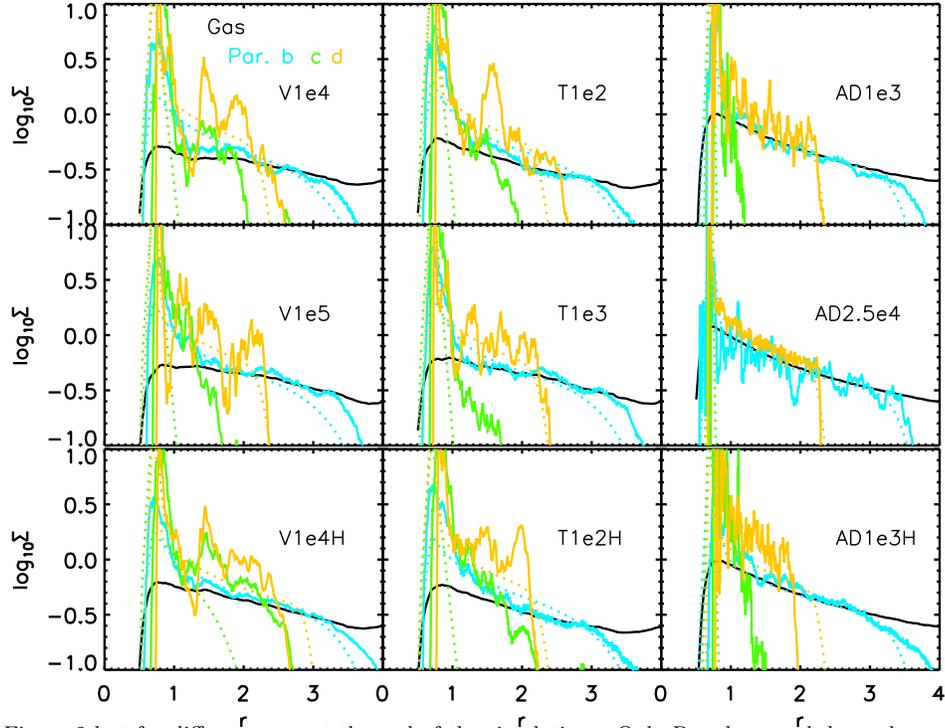} 
\vspace{-0.3 cm}
\caption{Similar to Figure \ref{fig:radialtorpaper1} but for
different runs at  the end of the simulations. Only Par. b, c, and d are shown.
 In each panel, the dotted curves
are from the 1-D model used in Figure
\ref{fig:radialtorpaper1}.  
} \label{fig:radialpaper}
\end{figure*}

\begin{figure*}[ht!]
\centering
\includegraphics[width=0.8\textwidth]{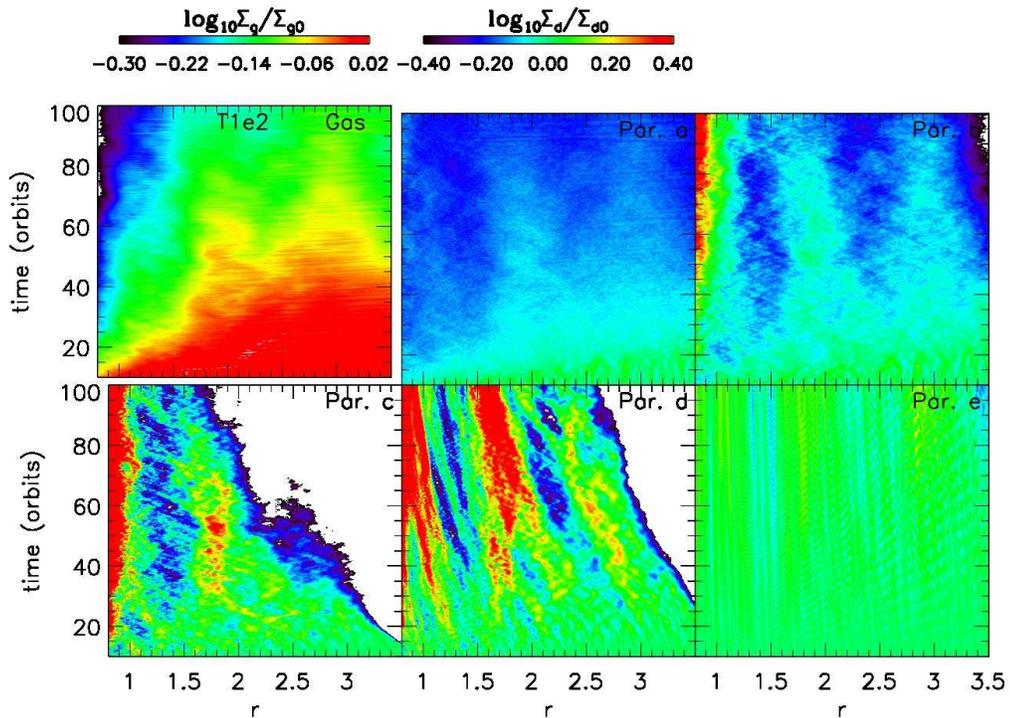} 
\vspace{-0.1 cm}
\caption{Space time diagram of the relative surface density in the radial direction for gas and various types of particles in run T1e2.
In the gas panel, besides the decrease of the surface density due to gas accretion, two density peaks  are visible at $r \sim 2$ and $r \sim 3$. Since these gas features can trap dust particles, the density peaks become 
more apparent in the Par. b panel,  most prominent in the Par. c and d panels, and also noticeable in the Par. e panel.
 } \label{fig:zonal}
\end{figure*}

\begin{figure*}[ht!]
\centering
\includegraphics[width=0.8\textwidth]{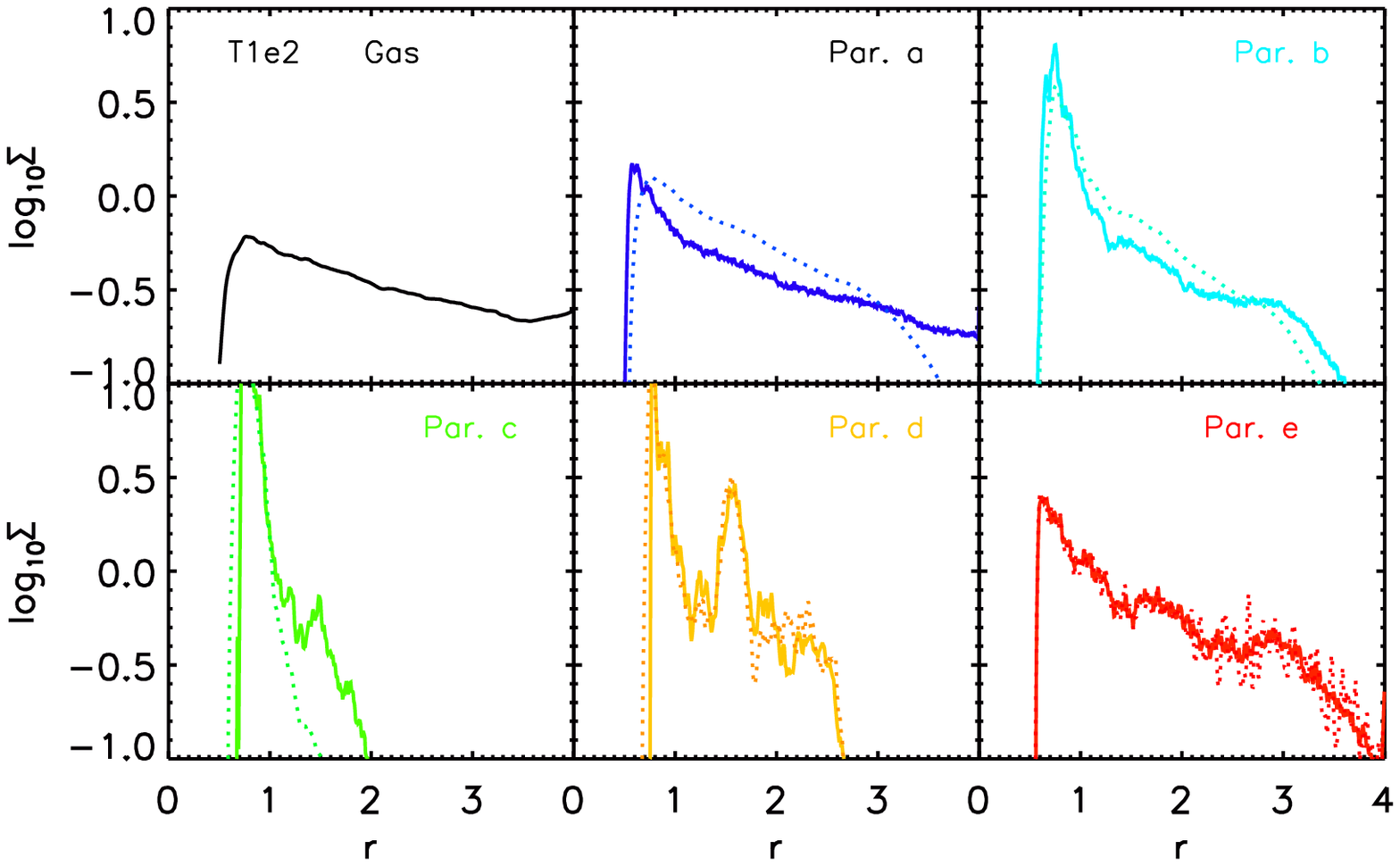} 
\vspace{-0.3 cm}
\caption{Similar to Figure \ref{fig:radialtorpaper1},  but the dotted curves are from the refined
1-D model with $\Sigma_{g}$ and $D_{d,r}=\alpha c_{s}^2/\Omega$
inputted from the MHD simulation. The  agreement between this refined 1-D model and the simulation is significantly
improved compared with Figure \ref{fig:radialtorpaper1}. In particular, particle trapping in the density peaks 
is very well reproduced (Par. d and e), suggesting the refined 1-D model is adequate to study particle trapping in zonal flows.
However, even this 1-D model could not fully reproduce the distribution of Par. c, which has $T_{s}\sim$1.
} \label{fig:radialtorpaper2}
\end{figure*}

\subsection{Evolution of Dust Surface Density}
{  In order to test if the evolution of dust in our 3-D turbulent simulations
can be reproduced by a 
 simple 1-D  dust model (Equation (\ref{eq:sigd})), we solve Equations (\ref{eq:sigd}) to 
(\ref{eq:driftv}) with the same disk structure in  3-D simulations.}
Equations (\ref{eq:sigd}) to 
(\ref{eq:driftv}) are solved 
using the operator split finite difference scheme similar to Stone \& Norman (1992).
The diffusion term in Equation (\ref{eq:diff}) is added using the central difference scheme, 
while the drift term in Equation (\ref{eq:driftv}) is 
added using the van Leer upwind method. At the radial boundaries, the dust density is set to be zero.
In order to solve the evolution of dust, we also need to specify how 
gas ($\Sigma_{g}(t)$) evolves with time in Equations (\ref{eq:diff}) and (\ref{eq:driftv}). 
To approximate the evolution of the gas disk, 
 two 1-D gas models have  been constructed using MHD simulations with different 
degrees of sophistication. 
The first model does not require knowing how the gas surface density 
fluctuates with time due to MRI turbulence. A fixed gas surface density has been applied.
In the second model, the gas surface density at every time step is  directly inputted from MHD simulations at that time.

In the first 1-D model,  based on the fact that the gas disk does not evolve much within 100 orbits,
we fix the gas surface density as
\begin{eqnarray}
\Sigma_{g}&=&\Sigma_{0}r^{-1}\,\,\,\,\,\,\,\,\,\,\,\,\,\,\,\,\,\,\,\,\,\,\,\,\,\,\,\,\, {\rm at}\,\,\,\, r\ge0.7 \nonumber\\
\Sigma_{g}&=&\Sigma_{0}r^{-1}{\rm exp}((r-0.7)/0.2)\,\,\,\,\, {\rm at}\,\,\, 0.5\le r\le0.7\, \label{eq:sigg0}
\end{eqnarray}
where $\Sigma_{0}=1$.
The initial dust surface density equals the gas surface density.
The exponential cut-off within $r=0.7$ is to mimic the gas surface density in MRI global
simulations where the inner disk within $r=0.7$ is depleted due to the inner open boundary condition.
$D_{d,r}$ is equal to $D_{d,x}$ in Equation (\ref{eq:ddx2}).
Since $t_{eddy}=\Omega^{-1}$ is a good 
assumption in ideal MHD (\S 5 and Carballido \etal 2011), $D_{g,x}$ is simplified to
$\langle v_{g,r}^{2}\rangle \Omega^{-1}$.  Furthermore, with
$\langle v_{g,r}^{2}\rangle\sim\alpha c_{s}^2$ and $\alpha\propto r^{-5/4}$ for our disk setup, we have $D_{g,x}\propto r^{-1/4}$.
Thus, we set $D_{g,x}=\langle v_{g,r}^{2}\rangle_{r=1} \Omega_{0}^{-1} (r/r_{0})^{-1/4}$, where $\langle v_{g,r}^{2}\rangle_{r=1}$
is given in Table 1. 

{ The comparison between this simple 1-D model and the 3-D turbulent simulation T1e2 at 100 orbits 
 is shown in Figure 
\ref{fig:radialtorpaper1}, which demonstrates that the main features in the 3-D simulation can be reproduced by
the 1-D model.} 
The solid curves in  Figure 
\ref{fig:radialtorpaper1} are the 
azimuthally averaged surface density for the gas (black curve) and different particle types (colored curves)
in the simulation, while the
dotted curves are the same quantities from the simple 1-D model. In the gas panel,  the upper black solid curve is the initial gas surface density in the simulation,
while the lower black solid curve is the gas surface density at 100 orbits. The shaded region is thus due to disk accretion, which confirms that the gas
disk does not evolve much during 100 orbits since this is significantly shorter than the viscous timescale. 
For most particle types, both the amplitude and size of the dust disks are similar between the 1-D model and
3-D simulations \footnote{The comparison
is not good for Par. a at the outer disk edge since the boundary conditions are different in 3-D simulations
and 1-D models. } 

{  However,
for Par. c and d having $T_{s}\sim 1$, there is a noticeable  discrepancy between the simulation and the 1-D model.} The 1-D model
predicts that all Par. c should have drifted to the inner boundary, while the simulation has a significantly extended disk. 
For Par. d, not only the position of the outer edge is different but also
dust in the MHD simulations show large amplitude density peaks. These peaks for Par. d are also present in other ideal MHD 
simulations (left two panels in Figure \ref{fig:radialpaper}) where the disk is very turbulent. 

{  These dust density peaks in simulations are related to the large-scale 
gas features induced by MRI turbulence (e.g., zonal flows ).}
Examining the gas surface density closely in Figure \ref{fig:radialpaper} (black solid curves),
we can see there are small amplitude density fluctuations in the gas disk. Although these fluctuations only change the gas surface
density slightly, they can affect dust significantly since the dust drift speed sensitively depends on the slope of the
gas surface density. The effect is that dust particles always drift to pressure peaks in disks. 
Figure  \ref{fig:zonal} shows the space time diagram for the disk surface density in run
T1e2. In order to illustrate the density fluctuations instead of the background density gradient,
we have divided the surface density at $t$ by the initial surface density. The top left panel shows the gas surface density.
Besides the general trend of the decreasing gas surface density due to accretion, 
we can see two density peaks at $r\sim 2$ and $3$ which
persist over the whole simulation. 
Since these gas features can trap dust particles, the density peaks become 
more apparent in the Par. b panel,  most prominent in the Par. d panel, and also noticeable in the Par. e panel.
More discussions on the amplitude of zonal flows in various simulations and how they trap particles are presented in \S 6.2.  

{ To take into account the gas fluctuations induced by MRI turbulence, 
we construct a second 1-D evolutionary model for the dust disk.}
In this model, we first output the azimuthally averaged gas surface density ($\Sigma_{g}$) and equivalent $\alpha$
 from MHD simulations at every 0.1 orbit. 
Then we solve Equations (\ref{eq:sigd}) to
(\ref{eq:driftv}) by inputing this time evolving 
$\Sigma_{g}$ and $D_{d,r}\sim \alpha c_{s}^{2}/\Omega$. The dust will then respond to the
evolving gas disk. Using this model for T1e2 run, we derive
the surface density for dust at 100 orbits, which is plotted as the dotted curves in Figure \ref{fig:radialtorpaper2}.
For Par. c, this model still produces a much smaller disk than the simulation, suggesting that
 1-D axisymmetrically averaged disk models
cannot reproduce the evolution of dust  with $T_{s}\sim 1$, and
non-axisymmetric density features in disks  (e.g., spiral arms) affect the evolution of these particles significantly.
However, this refined 1-D  model reproduces the surface density evolution for other particle types very well. Especially for
 Par. d and e, the 1-D model reproduces the dust
density peaks very accurately, confirming that
the density peaks in the dust are due to axisymmetric structures in the gas disk (e.g., zonal flows).  

In AD runs, since the turbulence is very weak, 
the evolution of the dust in our simulations is dominated by the radial drift. To study 
radial turbulent diffusion in AD cases, we need to carry out shearing box simulations 
where the dust radial drift  is zero. These simulations are carried out and studied in \S 5. 

\subsection{Dust Vertical Settling}

\begin{figure*}[ht!]
\centering
\includegraphics[width=0.8\textwidth]{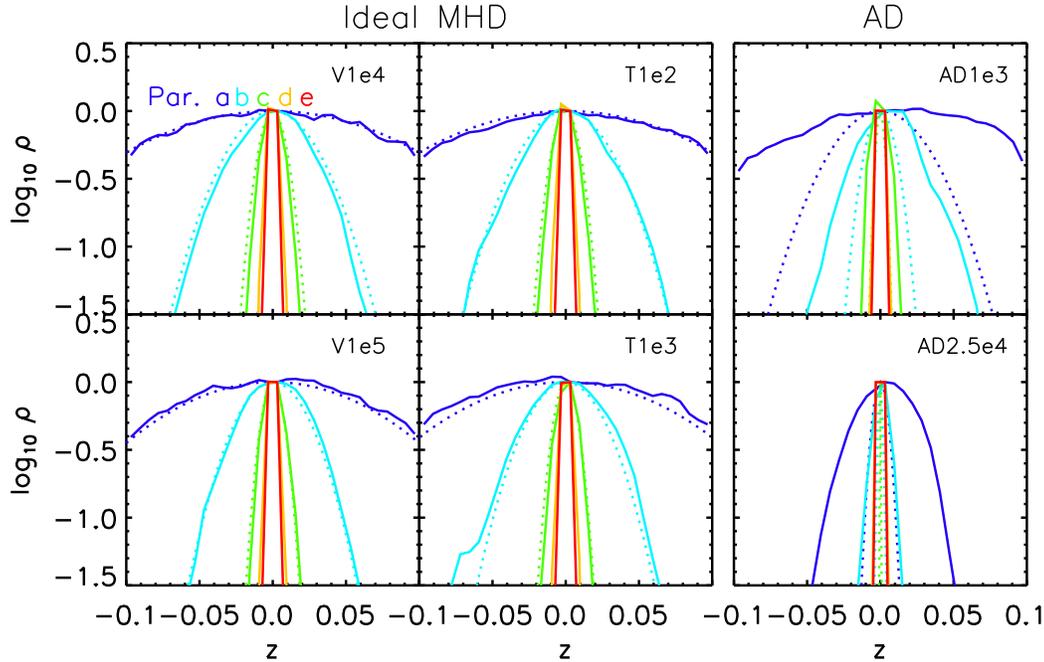} 
\vspace{-0.3 cm}
\caption{Azimuthally averaged density at $r=1$ with respect to
disk height for different particle types (colored curves) in
ideal and non-ideal MHD runs at 100 orbits. The dotted curves are the Gaussian profiles 
with $h_{d}$ (Equation (\ref{eq:eqhd3})) derived  with $t_{eddy}=(2\Omega)^{-1}$. 
The agreement is very good
for ideal MHD runs. For AD runs (the right panels), this $h_{d}$ with $t_{eddy}=(2\Omega)^{-1}$
significantly underestimates the dust scale height in simulations.
} \label{fig:vertical}
\end{figure*}

{ Besides the radial drift, particles will also settle to the disk midplane.}
In turbulent disks, a steady density profile in the $z$ direction is established 
when particle settling is balanced by turbulent diffusion, and the profile 
can be derived analytically as given by Equations (\ref{eq:hds3}) and (\ref{eq:eqhd3}).  

{ Thus, we compare the dust vertical structure in our simulations with these analytical
models in Figure \ref{fig:vertical}.} The solid curves are the vertical dust density profiles 
in our simulations at $r=1$, while the dotted curves have Gaussian density profiles 
with the scale height ($h_d$) from Equation (\ref{eq:eqhd3}). Although the simulations cannot resolve dust scale height of 
particles with $T_{s}>1$ (e.g. Par. d and e), we find 
Par. a to c in ideal MHD simulations can be well fitted by the analytical model assuming $t_{eddy}=(2\Omega)^{-1}$. 
This is consistent with previous MHD simulations that
$h_{d}$ with $t_{eddy}\sim \Omega^{-1}$ can reproduce the dust vertical structure in ideal MHD simulations.

{ However, when AD dominates in disks, this analytical profile with $t_{eddy}=(2\Omega)^{-1}$ significantly
underestimates the dust scale height in simulations by more than a factor of 2 (right panels in Figure \ref{fig:vertical}).} 
To explain these relatively thick dust disks, Equation (\ref{eq:eqhd3}) requires that
$t_{eddy}$ has to be several times of $\Omega^{-1}$ in AD runs which 
is much larger than $t_{eddy}\sim\Omega^{-1}$ in ideal MHD runs.
 
\section{Shearing Box Simulations}
\begin{table*}[ht]
\begin{center}
\caption{Shearing Box Simulations \label{tab2}}
\begin{tabular}{cccccccccccccccc}
\tableline
Run & Reso. & $X\times Y\times Z$ &$\beta_{0}$ &  $\langle\langle v_{g,x}^{2}\rangle\rangle$ & $\langle\langle v_{g,y}^{2}\rangle\rangle$ & $\langle\langle v_{g,z}^{2}\rangle\rangle$ & $\alpha_{R}$ \tablenotemark{a} &  $\alpha_{M}$\tablenotemark{b}  & $\alpha$ & $\frac{d \langle x^{2}\rangle}{2dt}$ & $\frac{d\langle z^{2}\rangle}{2dt}$ & $t_{eddy,x}$ & $t_{eddy,z}$ & $Sc_{x}$ & $Sc_{z}$\\
name &    & h$\times$h$\times$h &  & ($c_{s}^{2}$) &  ($c_{s}^{2}$) & ($c_{s}^{2}$) &  &    & & ($H^{2}\Omega$) & ($H^{2}\Omega$)    & $\Omega^{-1}$& $\Omega^{-1}$      & &\\
\tableline
\tableline
Unstratified\\
\tableline
VR32 & 32/h & 1$\times$4$\times$2 & Vert. 1e4 & 0.050 & 0.046 & 0.021 & 0.016 & 0.072  & 0.088 & 0.073 & 9.9e-3 & 1.5 & 0.47 &1.2 & 8.9\\
TR32 & 32/h &  1$\times$4$\times$2 & Tor. 1e3 & 0.021 & 0.015 & 0.010 & 6.4e-3 & 0.025  & 0.031 & 0.020 & 9.2e-3 & 0.95 & 0.92  & 1.5 & 3.4\\
ZR32 & 32/h &  1$\times$4$\times$2 & Zero 1e4 & 0.013 & 8.1e-3 & 5.8e-3 & 3.8e-3 & 0.014  & 0.018 & 0.014 & 4.5e-3 & 1.1 & 0.78 & 1.3 & 4.0 \\
ZR64 & 64/h &  1$\times$4$\times$2 & Zero 1e4 & 6.3e-3 & 4.5e-3 & 3.0e-3 & 1.7e-3 & 6.8e-3  & 8.5e-3 & 6.4e-3 & 2.8e-3 & 1.0 & 0.93 & 1.3 & 3.0 \\
\tableline
AD1R32 & 32/h &  1$\times$4$\times$2 & Vert. 1e3 & 5.7e-3 & 1.2e-3 & 4.3e-3 & 1.1e-3 & 2.9e-3  & 3.9e-3 & 0.026 & 0.013 & 4.6& 3.0 & 0.15 & 0.3 \\
AD1R64 & 64/h &   1$\times$4$\times$2 &Vert. 1.e3 & 5.4e-3 & 1.3e-3 & 4.2e-3 & 1.0e-3 & 2.5e-3  & 3.6e-3 & 0.015 & 0.010 & 2.8 & 2.4 & 0.24 & 0.36 \\
AD2R32 & 32/h &  1$\times$4$\times$2 & Vert. 1e4 & 8.8e-4 & 1.5e-4 & 2.7e-4 & 1.4e-4 & 3.5e-4  & 4.9e-4 & 9.3e-4 & 1.1e-3 & 1.1 & 4.1 & 0.5 & 0.45\\
AD2R64 & 64/h &  1$\times$4$\times$2 & Vert. 1e4 & 2.9e-3 & 5.e-4 & 4.4e-4 & 5.2e-4 & 4.4e-4  & 9.5e-4 & 1.3e-3 & 1.4e-3 & 0.45 & 3.2 &0.73 & 0.68\\
AD2R128 & 128/h &  1$\times$4$\times$2 & Vert. 1e4 & 1.6e-3 & 3.4e-4 & 5.4e-4 & 2.4e-4 & 4.2e-4  & 6.6e-4 & 9.3e-4 & 1.4e-3 & 0.58 & 2.6 & 0.71& 0.47 \\
AD2R32W & 32/h &  8$\times$4$\times$2 &Vert. 1e4 & 3.8e-3 & 4.1e-4 & 2.8e-4 & 5.1e-4 & 4.0e-4  & 9.0e-4 & 1.4e-3 & 8.9e-4 & 0.37 & 3.2 & 0.64& 1.0 \\
AD3R32 & 32/h &  1$\times$4$\times$2 & Vert. 2.5e4 & 6.8e-4 & 8.6e-5 & 8.2e-5 & 9.4e-5 & 1.4e-4  & 2.4e-4 & 4.0e-4 & 3.4e-4 & 0.59 & 4.1 & 0.60 &  0.71\\
AD3R64 & 64/h &  1$\times$4$\times$2 & Vert. 2.5e4 & 1.9e-3 & 3.1e-4 & 1.9e-4 & 3.3e-4 & 1.7e-4  & 5.0e-4 & 4.6e-4 & 6.0e-4 & 0.24 & 3.2 & 1.1 &  0.83\\
\tableline
\tableline
Stratified\\
\tableline
VSR32 & 32/h &  4$\times$4$\times$8 & Vert. 1e4  \\
\tableline  
z=[-3h,-h]\tablenotemark{c} & -& -& -& 0.10 & 0.069 & 0.063 & 0.029 & 0.16  & 0.19 & 0.12 & 0.012 & 1.2 & 0.19 & 1.6 & 16 \\
z=[-h,h] & -& -& -& 0.052 & 0.042 & 0.035 & 0.016 & 0.066  & 0.082 & 0.059 & 4.2e-3 & 1.1& 0.12 & 1.4 & 20\\
z=[h,3h]\tablenotemark{c}  & -& -& -& 0.12 & 0.086 & 0.087 & 0.033 & 0.18  & 0.21 & 0.078 & 3.6e-3 & 0.65 & 0.041 & 2.7& 58\\
\tableline
AD2SR32 & 32/h & 4$\times$4$\times$6 & Vert. 1e4  \\
\tableline  
z=[-3h,-h] & -& -& -&3.8e-3 & 7.3e-4 & 2.8e-3 & 5.3e-4 & 8.2e-4  & 1.4e-3 & 1.1e-3 & 2.2e-3 & 0.29 & 0.79 & 1.3 & 0.64\\
z=[-h,h] & -& -& -& 1.1e-3 & 2.2e-4 & 8.8e-4 & 1.7e-4 & 5.4e-4  & 7.1e-4 & 9.8e-4 & 1.9e-3 & 0.89& 2.2 & 0.72 & 0.37\\
z=[h,3h] & -& -& -& 3.0e-3 & 3.8e-4 & 2.2e-3 & 3.8e-4 & 8.8e-4  & 1.3e-3 & 1.6e-3 & 2.8e-3 & 0.53 & 1.3 & 0.81& 0.46\\
\tableline
AD2SLR32 & 32/h & 4$\times$4$\times$8  &Vert. 1e4  \\
\tableline  
z=[-3h,-h] & -& -& -& 6.9e-3 & 2.8e-3 & 5.3e-3 & 2.7e-3 & 2.9e-2  & 3.1e-2 & 5.3e-4 & 1.3e-3 & 0.077 & 0.25 & 58 & 24\\
z=[-h,h] & -& -& -&5.3e-4 & 1.3e-4 & 7.5e-4 & 9.5e-5 & 1.1e-3  & 1.2e-3 & 1.1e-3 & 2.6e-3 & 2.1& 3.5 & 1.1 & 0.46\\
z=[h,3h] & -& -& -& 3.3e-3 & 2.9e-4 & 3.8e-3 & 1.4e-3 & 3.6e-2  & 3.8e-3 & 6.3e-3 & 2.1e-3 & 1.9 & 0.55 & 0.60& 1.8\\
\tableline
\end{tabular}
\tablenotetext{1}{$\alpha_{R}\equiv\langle\langle \rho v_{g,x} \delta v_{g,y} \rangle\rangle/\langle\langle \rho \rangle\rangle c_{s}^2$}
\tablenotetext{2}{$\alpha_{M}\equiv\langle\langle -B_{x}B_{y}\rangle\rangle/\langle\langle \rho \rangle\rangle c_{s}^2$}
\tablenotetext{3}{The diffusion coefficients are measured by tracing particles from 30 to 33 orbits.}
\end{center}
\end{table*}

It is surprising that the precense of AD significantly affects $t_{eddy}$ and $D_{g,z}$.
To verify this result and unveil the physics behind it,
we have carried out both unstratified and stratified shearing box simulations including Lagrangian dust particles.

\subsection{Unstratified Simulations}
{  The unstratified simulations have a similar setup as Bai \& Stone (2011).} For most runs, the box size  is
$[-0.5 h,0.5 h]\times[-2 h, 2 h]\times[-h, h]$ in the $x$, $y$, and $z$ directions. For ideal MHD simulations, disks are threaded by
net vertical, net
toroidal, or zero net fields. 
For simulations with AD, Am is again chosen as one and only net vertical fields have been applied.
The detailed simulation
parameters are shown in Table 3. The nomenclature for run names in Table 3 is as following. 
In ideal MHD runs, ``V'', ``T'', and ``Z'' denote that the disk is threaded by net vertical, toroidal, or zero magnetic flux.
In AD runs, ``AD1'', ``AD2'', and ``AD3''  represent the disk with initial 
plasma $\beta$ of 10$^{3}$, 10$^{4}$, and 2.5$\times$10$^{4}$. ``R32'', ``R64'', and ``R128'' 
indicate that the numerical resolution is $32/h$, $64/h$, and $128/h$.

In these simulations, there are seven types of particles with stopping times of
$T_{s}=\{10^{-3}, 10^{-2}, 10^{-1}, 1, 10, 10^{2}, 10^{3}\}$, which are denoted as
Par. A to G (different from Par. a to g in global simulations). For each type of particles, we have 
put 10$^{4}$ particles uniformly in the box. We have run these simulations to 100 orbits. Although the gas
is unstratified, particles feel the gravitational force to the disk midplane similar to our global 
unstratified simulations.

{ Turbulence in these simulations has similar properties as in global unstratified simulations in \S 4.}
The averaged mean squared velocities, stresses, and equivalent 
$\alpha$ are given in Table 3. The average is taken from the whole simulation box and over the time from snapshots
at each orbit from 60 to 100 orbits. 
Similar to the global runs, turbulence is significantly suppressed by AD,
and all the ideal MHD simulations have
$ \alpha c_{s}^{2} \sim \langle\langle v_{g,x}^2\rangle\rangle > \langle\langle v_{g,y}^2\rangle > \langle\langle v_{g,z}^2\rangle\rangle $, 
while disks dominated by AD have $ \langle\langle v_{g,x}^2\rangle\rangle \gtrsim  \langle\langle v_{g,z}^2\rangle\rangle$ 
and $\langle\langle v_{g,x}^2\rangle\rangle > \alpha c_{s}^2$. 

\begin{figure*}[ht!]
\centering
\includegraphics[width=0.76\textwidth]{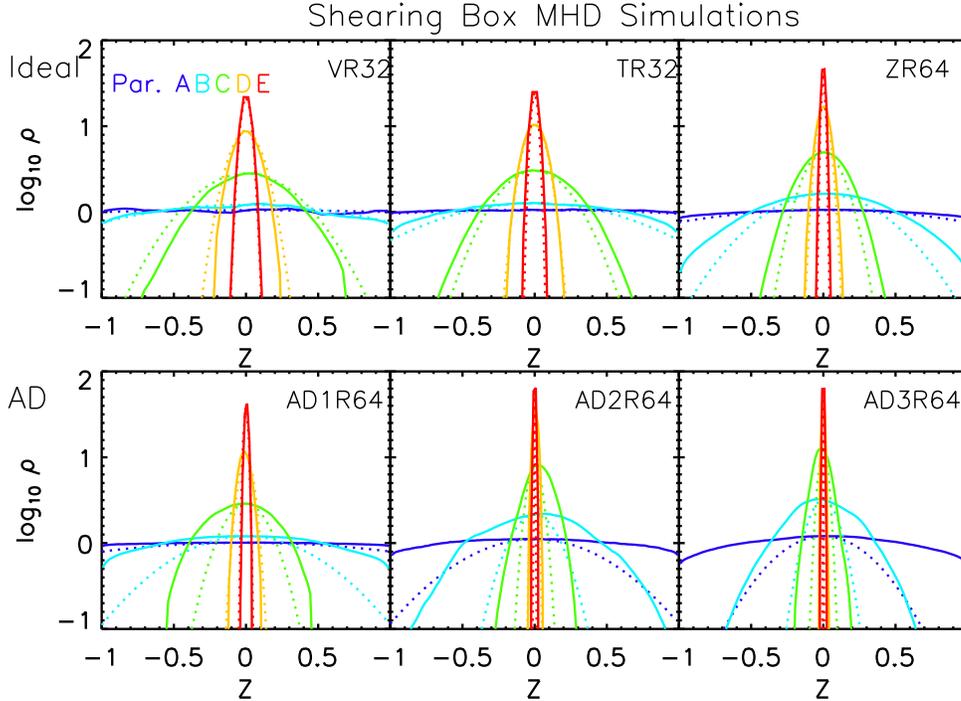} 
\vspace{-0.2 cm}
\caption{Space and time averaged density with respect to
the disk height for different types of particles (colored curves) in shearing box simulations in
 ideal MHD (upper panels) and non-ideal MHD with AD (lower panels). 
 The dotted curves are from the Gaussian profiles 
using $h_{d}$ (Equation (\ref{eq:eqhd3}))  with $t_{eddy,z}=(2\Omega)^{-1}$. 
As in Figure \ref{fig:vertical}, 
the agreement is only good
for ideal MHD runs. 
} \label{fig:verticalShear1}
\end{figure*}

{ These simulations have also confirmed that $h_{d}$ with $t_{eddy}\sim \Omega^{-1}$ significantly underestimates 
the dust scale height in disks dominated by AD.} 
Figure 
\ref{fig:verticalShear1} shows  the dust vertical density profiles  after being averaged at every orbit from 60 to 100 orbits.
The profiles confirm the results found in our global simulations (e.g., Figure \ref{fig:vertical})
that $h_{d}$ with $t_{eddy,z}=(2\Omega)^{-1}$ (Equation (\ref{eq:eqhd3})) can fit the dust density 
profiles in ideal MHD runs well (at least for Par. A to C with a resolved scale height), but
a larger $h_{d}$---thus a larger $D_{g,z}$ and $t_{eddy,z}$---is needed to explain density profiles in simulations dominated by AD.

{ We have traced the trajectory of 
each individual particle in shearing box simulations, so that
we can measure the dust diffusion coefficients directly using Equations (\ref{eq:ddz}) and (\ref{eq:ddx}).} For each type of particles,
we randomly pick 160 particles and measure their rates of change of $x^{2}$ and $z^{2}$ from 30 to 50 orbits.
Then we average these rates to derive the diffusion coefficients in both radial and vertical directions.
Our smallest particles couple
with the gas so well that we use their $D_{d}$ to represent the gas diffusion coefficients $D_{g}$. Both $D_{g,x}$ and $D_{g,z}$
are given in Table 3 for all our runs. 
$t_{eddy,x}$ and $t_{eddy,z}$ calculated by $D_{g,x}/\langle\langle v_{x}^2\rangle\rangle$
and $D_{g,z}/\langle\langle v_{z}^2\rangle\rangle$ are also given in Table 3.
All the ideal MHD runs 
have  $D_{g,x}\sim \langle\langle v_{x}^2\rangle\rangle \Omega^{-1}$ and 
$D_{g,z}\sim \langle\langle v_{z}^2\rangle\rangle \Omega^{-1}$, suggesting $t_{eddy}\sim \Omega^{-1}$ for both $v_{x}$ and $v_{z}$. 
This is again consistent with previous simulations (Johansen \& Klahr 2005).

{ However, in all our AD runs,  the measured $D_{g,z}$ is three to four times of $\langle\langle v_{z}^2\rangle\rangle \Omega^{-1}$,
so that $t_{eddy,z}\sim 3-4\Omega^{-1}$ which
is significantly
longer than the eddy time in ideal MHD ($t_{eddy,z}\sim \Omega^{-1}$).}
It is this large $t_{eddy,z}$ that causes the relatively large dust scale height  in both global
and local AD simulations. Another anomaly in AD runs is that $t_{eddy,x}\neq t_{eddy,z}$.
$t_{eddy,x}$ can be
5 times larger than $\Omega^{-1}$ (AD1 with $\beta_{0}=10^{3}$), 
comparable to $\Omega^{-1}$ (AD2 with $\beta_{0}=10^{4}$)), or even 4 times smaller than 
$\Omega^{-1}$ (AD3 with $\beta_{0}=2.5\times10^{4}$)). 

\begin{figure*}[ht!]
\centering
\includegraphics[width=0.76\textwidth]{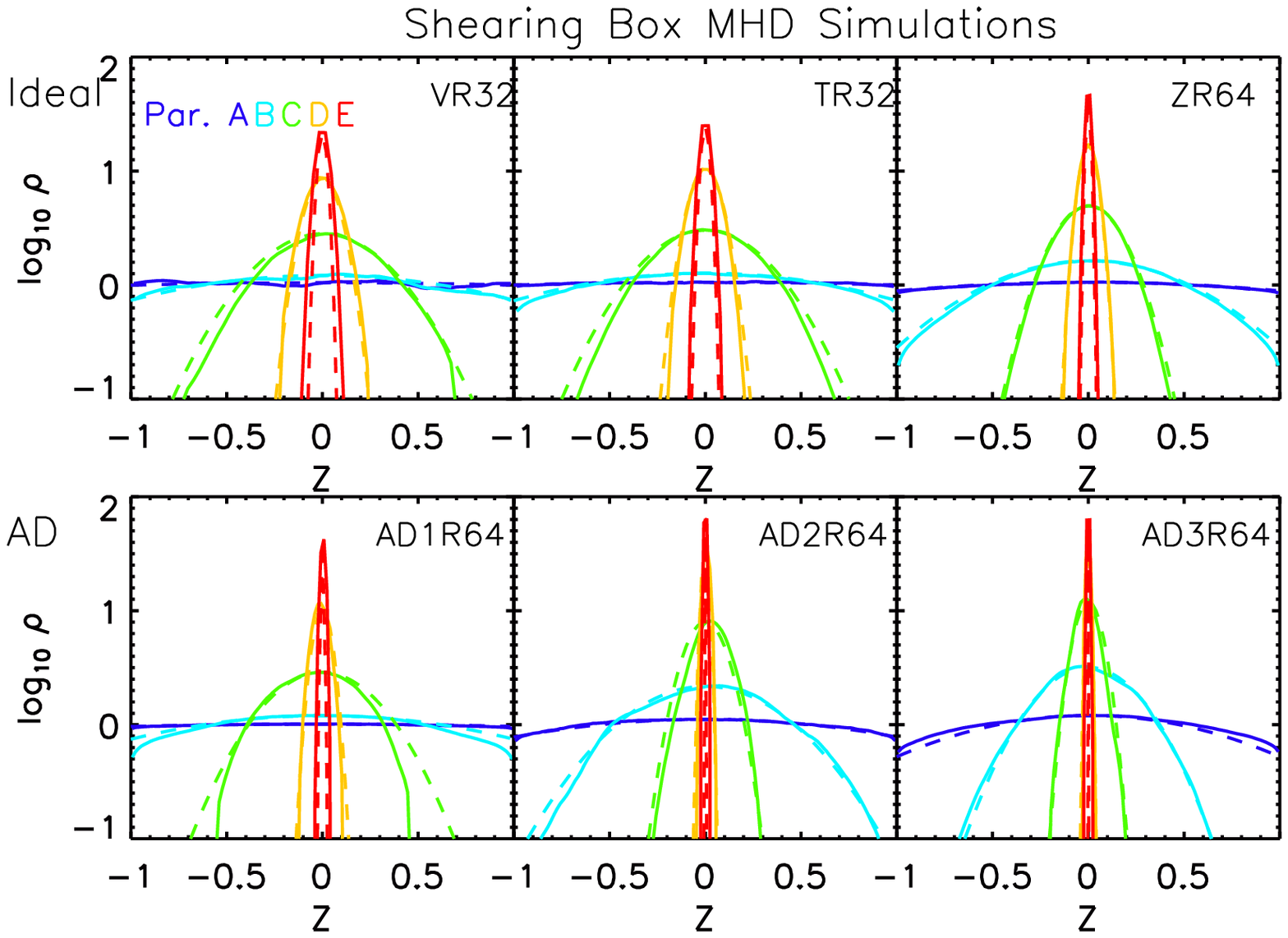} 
\vspace{-0.2 cm}
\caption{Similar to Figure \ref{fig:verticalShear1}.  The dashed curves are from 
the Gaussian profiles 
using $h_{d}$ (Equation (\ref{eq:eqhd2}))  with the directly measured $t_{eddy,z}$ in Table 3. The agreement
is good for all the runs.} \label{fig:verticalShear2}
\end{figure*}

{ If we plug the measured $D_{g,z}$ or $t_{eddy,z}$ from Table 3 into Equation (\ref{eq:eqhd2})
to derive the analytical
density profiles,
the resulting Gaussian profiles can fit the simulation results in both ideal MHD and AD runs very well.}
These density profiles using $D_{g,z}$ or $t_{z, eddy}$
in Table 3 are shown in Figure \ref{fig:verticalShear2} as the dashed curves, compared with
the simulation results as the solid curves. This confirms that the large $D_{g,z}$ and $t_{z,eddy}$ are responsible for the thick dust disks
in AD runs.

\begin{figure*}[ht!]
\centering
\includegraphics[width=0.8\textwidth]{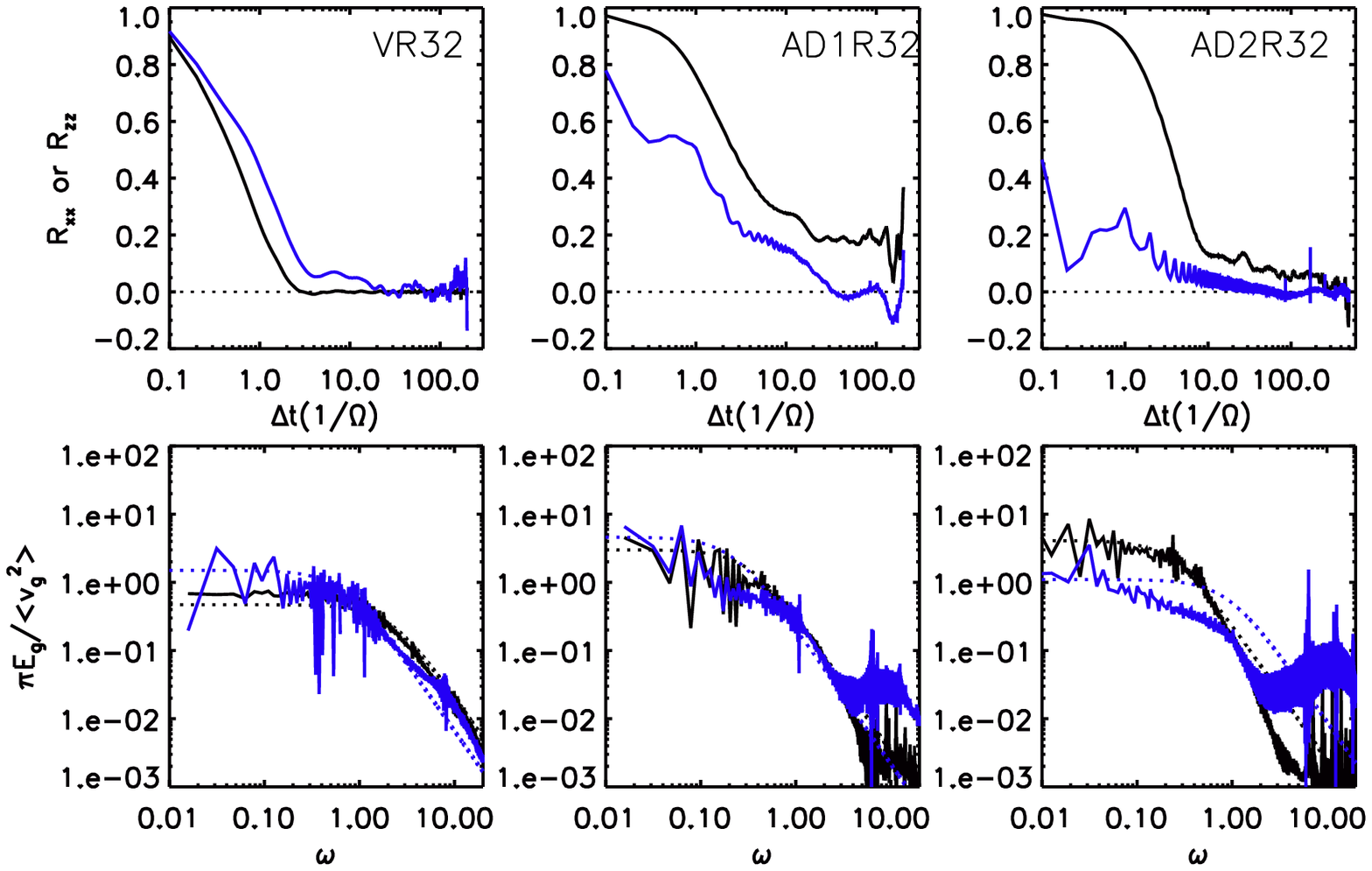} 
\vspace{-0.3 cm}
\caption{Upper panels:  the auto-correlation function $R_{xx}(\tau)/R_{xx}(0)$ (blue curves), $R_{zz}(\tau)/R_{zz}(0)$ (black curves) for
runs VR32, AD1R32, and AD2R32 (the left to right panels). AD runs have long correlation times for $v_{z}$.
Lower panels: The power spectrum for $v_{x}$ (blue curves) and  $v_{z}$ (black curves) 
for the same runs. The dotted curves are the analytic models from Equation (\ref{eq:ego}) using $t_{eddy,x}$ and $t_{eddy,z}$
in Table 3.  } \label{fig:autopower}
\end{figure*}

{ The anomaly of $t_{eddy}$ in AD runs
indicates that the presence of AD dramatically changes the properties of MRI turbulence, since 
$t_{eddy}$ is closely related to the turbulence autocorrelation function and the power spectrum as in
Equations (\ref{eq:dgzf}), (\ref{eq:Eg0}), and (\ref{eq:dgz}).} Figure \ref{fig:autopower}
shows the auto-correlation function $R(\tau)=\langle v_{g}(\tau)v_{g}(0)\rangle$ and the power spectrum 
for $v_{g,x}$ and $v_{g,z}$ in both ideal MHD
and AD runs. To derive the auto-correlation function and the power spectrum, we 
first output the simulation data (e.g., $v_{g,x}$ and $v_{g,z}$)
 using a time interval of $0.1 \Omega^{-1}$. 
We then shift the data along the $y$ 
direction a distance of $1.5\Omega x t$ to correct for the Keplerian shear. After the shift, we multiply  the velocity $v_{g}$ at each
time frame $t$ with the velocity at a later 
time $t+\tau$. Finally, $R(\tau)$ is derived by averaging  
$v_{g}(t+\tau)v_{g}(t)$ over both space and time.
The autocorrelation functions for both vertical and radial velocity, i.e., $R_{zz}(\tau)/R_{zz}(0)$ 
and $R_{xx}(\tau)/R_{xx}(0)$, in both ideal MHD
(VR32) and AD runs (AD1R32 and AD2R32) are shown in the upper panels of Figure \ref{fig:autopower}. 

{ The figure shows that although $R_{zz}$ in ideal MHD (VR32) drops to zero on the timescale of $\tau\sim\Omega^{-1}$,
$R_{zz}$ in AD runs (AD1R32 and AD2R32) drops off much more slowly on a timescale of $\tau\sim10 \Omega^{-1}$.}
Furthermore, in these AD runs, $R_{zz}$ still has a positive value  even at $\tau=100 \Omega^{-1}$.
Since
$t_{eddy,z}=\int_{0}^{\infty} R_{zz}(\tau)/R_{zz}(0) d\tau$,  the slow drop-off and positive tail of the
correlation function in AD runs lead to the large eddy time for $v_{z}$, consistent with the direct measurement of $t_{eddy,z}$ above through tracing dust particles.  

{ The autocorrelation function also reveals the anomalous turbulent diffusion in the $x$ direction in AD runs.}
Although $R_{xx}$ and $R_{zz}$ are almost identical in ideal MHD runs, 
they have very different profiles in AD runs.
In AD2R32, $R_{xx}$  is much smaller than $R_{zz}$, implying $t_{eddy,x}<t_{eddy,z}$, while $R_{xx}$ is 
close to $R_{zz}$ in AD1R32 implying $t_{eddy,x}\sim t_{eddy,z}$. These relationships are consistent with the values
measured in Table 3.

{ By doing Fourier transform for the autocorrelation functions 
(Equation (\ref{eq:egw})),
we also compute the power spectrum of turbulence ($\hat{E}_{g}(\omega)$) in both ideal and non-ideal MHD, shown in  the bottom panels of Figure \ref{fig:autopower}.}
We can see that, as $\omega \to 0$, $\pi\hat{E}_{g,z}(\omega)/\langle v_{g,z}^2\rangle$ in AD runs becomes $\sim 10 \Omega^{-1}$  which is significantly larger than $\sim \Omega^{-1}$
in the ideal MHD run. Since $t_{eddy,z}=\pi \hat{E}_{g,z}(0)/\langle v_{g,z}^2\rangle$ (Equation (\ref{eq:dgz})), this
 again implies a bigger $t_{eddy,z}$ in AD runs. 

\begin{figure}[ht!]
\centering
\includegraphics[trim=1cm 2cm 1cm 2cm, width=0.5\textwidth]{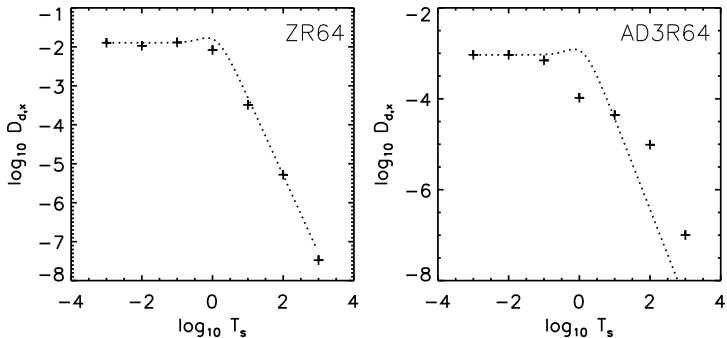} 
\vspace{-0.3 cm}
\caption{Particle radial diffusion coefficients as a function of particle stopping time, for ZR64 (the left panel) and AD3R64
(the right panel). The analytical model of YL (Equation (\ref{eq:ddx2})) is shown as dotted curves which agree with ideal MHD
simulations very well but not completely agree with AD runs.} \label{fig:diffmulti}
\end{figure}

{ For big particles with $T_{s}>1$, not only $\hat{E}_{g}(0)$ (or $t_{eddy}$) but also the whole $\hat{E}_{g}(\omega)$ over $\omega$ determines
the dust diffusion coefficients $D_{d,x}$ and $D_{d,z}$.} 
This is because, to derive $D_{d,x}$ and $h_{d}$ in Equations (\ref{eq:ddx1}) and (\ref{eq:eqhd2}) for big particles, we need to
 integrate the velocity and the power spectrum ($\hat{E}_{g}$, Equation (\ref{eq:ego})) over $\omega$.
This is different from small particles ($T_{s}<1$) which have the
 particle diffusion coefficient ($D_{d}$) equal to
the gas diffusion coefficient ($D_{g}$) that only depends
on the power spectrum $\hat{E}_{g}(\omega)$ at $\omega=0$ (Equation (\ref{eq:dgz})).
Thus, we compare the analytical power spectrum of Equation (\ref{eq:ego}) used in  Youdin \& Lithwick (2007) with the power spectrum derived from our simulations in
 the bottom panels of Figure  \ref{fig:autopower}. 
$t_{eddy,x}$ and $t_{eddy,z}$ in  Equation (\ref{eq:ego}) are given in Table 3. As shown, this  
analytical power spectrum (dotted curves) agrees well with the power spectrum in ideal MHD simulations.
However, it deviates from the power spectrum in AD runs significantly. The turnover between the integral scale and the inertial range, and
the slope in the inertial range are all different from the simple analytical formula. Moreover, the power spectrum in AD 
simulations
actually increases at the shortest timescales (largest $\omega$).

\begin{figure*}[ht!]
\centering
\includegraphics[width=0.8\textwidth]{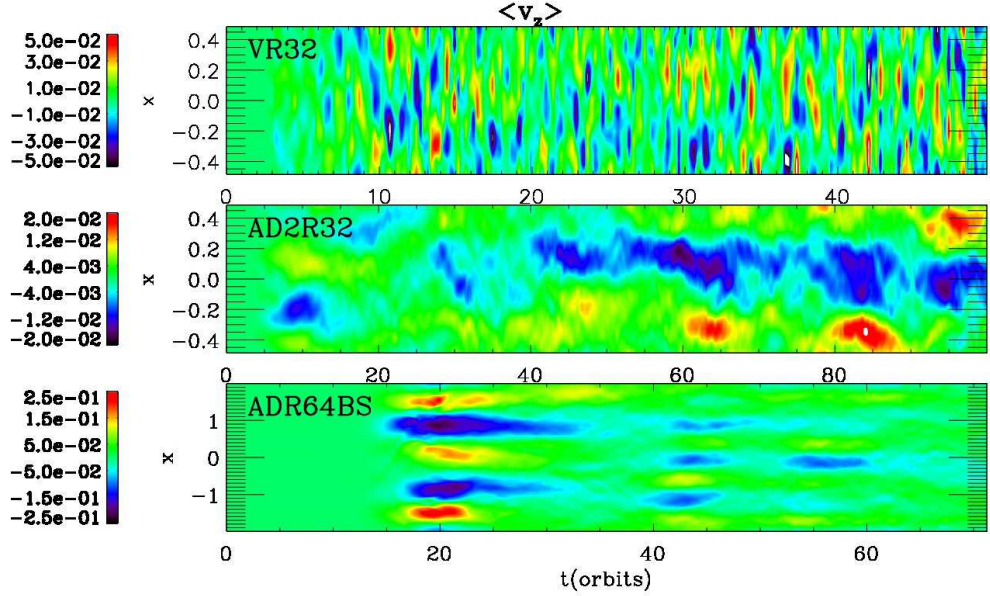} 
\vspace{-2.3 cm}
\caption{ Space time diagram of $\langle v_{z} \rangle$ in the $x$ direction for various runs.
$\langle v_{z} \rangle$ is derived by averaging $v_{z}$ in both $y$ and $z$ directions. 
In the ideal MHD run VR32, $v_{z}$ has a correlation time $\sim \Omega^{-1}$.
In AD runs,
the coherent $v_{z}$ forms horizontal bands in the diagram. 
} \label{fig:spacetime}
\end{figure*}

\begin{figure*}[ht!]
\centering
\includegraphics[width=0.8\textwidth]{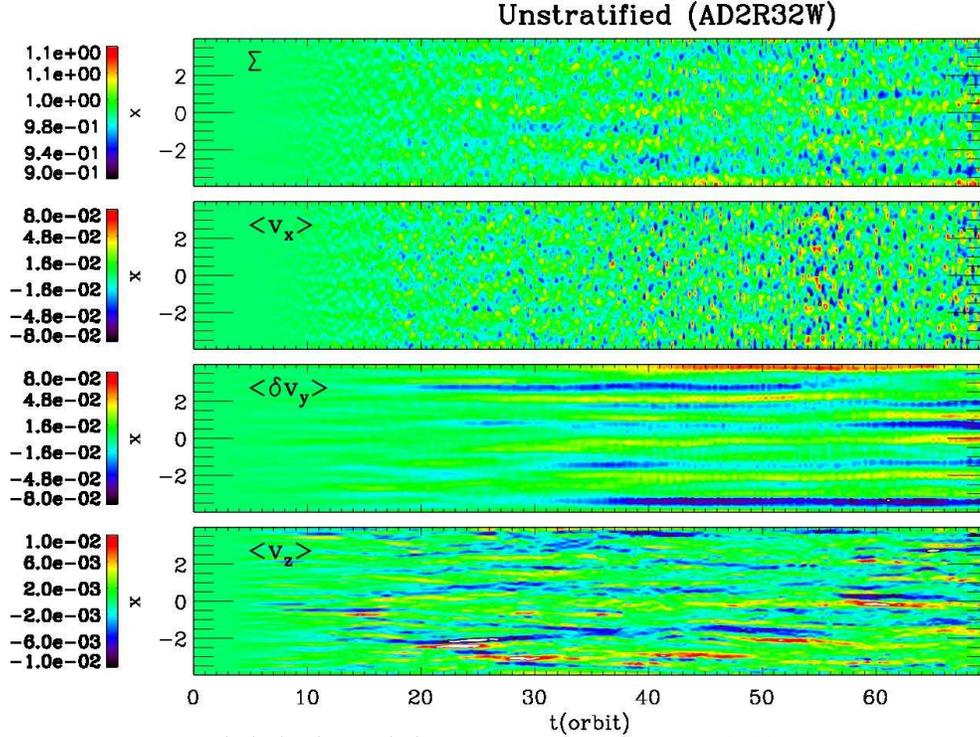} 
\vspace{-0.3 cm}
\caption{ Space time diagram of $\Sigma$, $\langle v_{x} \rangle$, $\langle \delta v_{y} \rangle$,
and $\langle v_{z} \rangle$ in the $x$ direction for run
AD2R32W. The weak zonal flows shown in the $\Sigma$ panel are more apparent in the
$\langle \delta v_{y} \rangle$ panel. $\langle v_{z}\rangle$ has a longer correlation time than $\langle v_{x}\rangle$.  
} \label{fig:unstratified}
\end{figure*}

{  Thus, this non-standard power spectrum in AD dominated disks implies
that  Equation (\ref{eq:eqhd2}) for $h_{d}$ and  Equation (\ref{eq:ddx1}) for $D_{d,x}$ 
derived using the standard power spectrum (Equation (\ref{eq:ego})) 
are not applicable for describing  diffusion of big particles in AD disks.}
To illustrate this point, Figure \ref{fig:diffmulti}
 compares the dust radial diffusion coefficients measured from simulations with the analytical
expression of Equation (\ref{eq:ddx2}). As expected, the analytical theory  reproduces
the radial diffusion coefficients in ideal MHD runs, but fails to reproduce the coefficients at
$T_{s}>1$ in AD runs. Unfortunately, we cannot test $h_{d}$ (Equation (\ref{eq:eqhd2})) for big particles with $T_{s}>1$, since it requires a much higher vertical resolution in simulations to resolve the scale height for these particles.

{ The long eddy time of $v_{z}$ in AD runs  should also manifest itself in physical space.}
We average $v_{z}$ in both $y$ and $z$ directions to derive $\langle v_{z} \rangle$, and plot
its space time diagram along the $x$ direction in Figure \ref{fig:spacetime}.
We also include one simulation in Bai \& Stone (2011) (denoted as ADR64BS) which
has Am=1 and $\beta_{0}$=1.e3, similar to our AD1R64 run.
In the ideal MHD run (VR32), $v_{z}$ has a coherence time $\sim \Omega^{-1}$. But in AD runs, 
coherent structure in $v_{z}$ 
can exist  for 100 orbits with a typical width of $h$ in the $x$ direction. 
On the other hand, Table 3
and Figure \ref{fig:autopower} suggest that, in AD runs,   $v_{x}$ should have a different
 coherent time than $v_{z}$. 
The space time diagrams for $\Sigma$, $\langle v_{x} \rangle$,
$\langle \delta v_{y} \rangle$, and $\langle v_{z} \rangle$ of AD2R32W are shown in Figure
\ref{fig:unstratified}. $\delta v_{y}$ is $v_{y}-v_{kep}$ where $v_{kep}=-1.5\Omega x$. The short correlation time of $v_{x}$ and the long correlation time of $v_{z}$ are apparent
in this figure, which is consistent with $t_{eddy,x}<t_{eddy,z}$ for this run in Table 3. 

 {The long-lived structure of $v_{z}$ in AD runs shown in Figure \ref{fig:spacetime} and \ref{fig:unstratified}
  implies that the linear growing mode still persists in the nonlinear phase, and such mode significantly affects particle dynamics. }
But the detailed mechanism deserves
further studies in future.

\subsection{Stratified Simulations}

\begin{figure*}[ht!]
\centering
\includegraphics[width=0.8\textwidth]{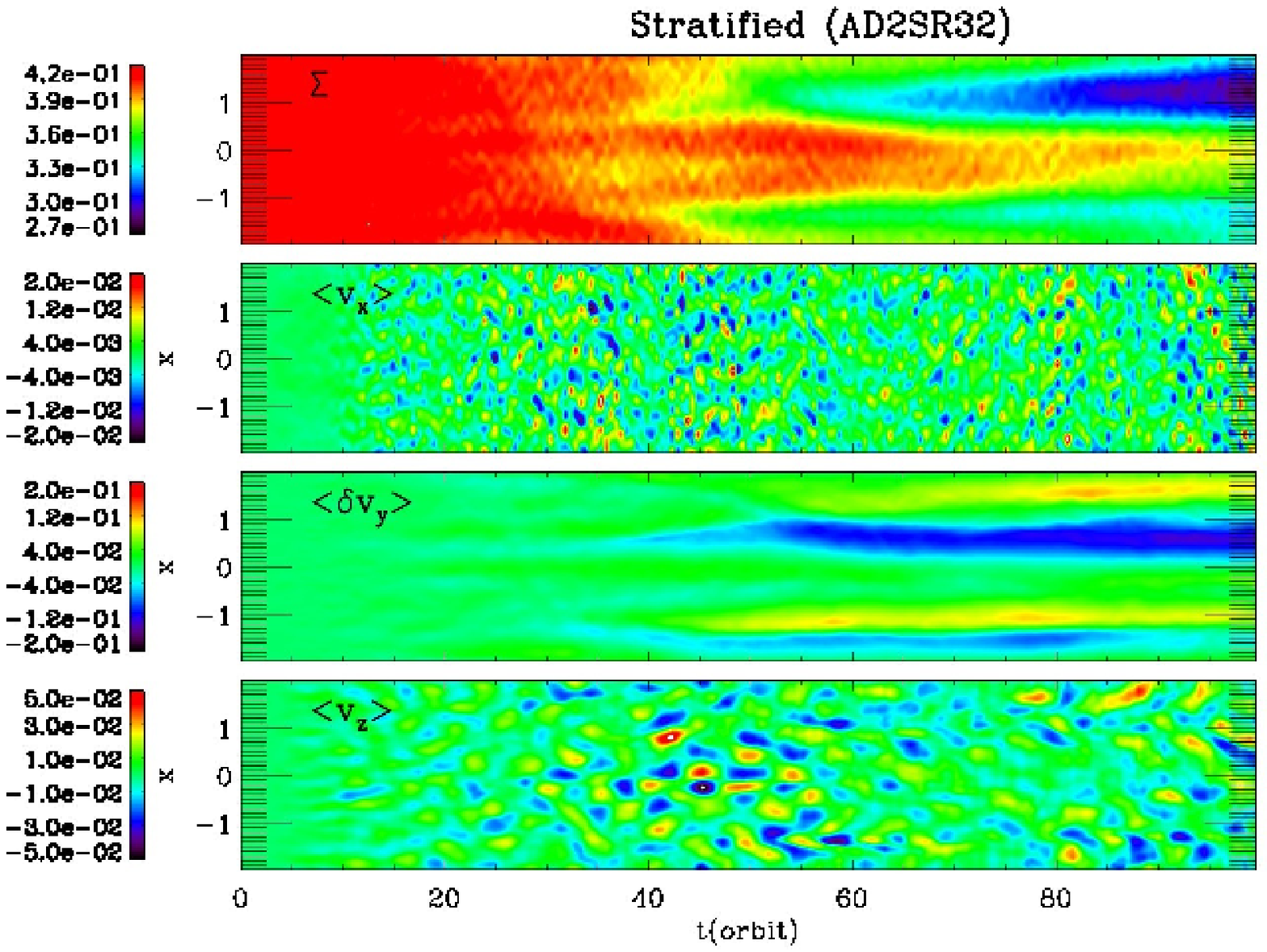} 
\vspace{-0.3 cm}
\caption{Similar to Figure \ref{fig:unstratified} but for AD2SR32. While
$v_{z}$ does not have a coherent value during 100 orbits in contrast to unstratified simulations, it still has a longer correlation time than $v_{x}$, implying
$t_{eddy,z}>t_{eddy,x}$.  } \label{fig:stratified}
\end{figure*}

\begin{figure}[ht!]
\centering
\includegraphics[width=0.5\textwidth]{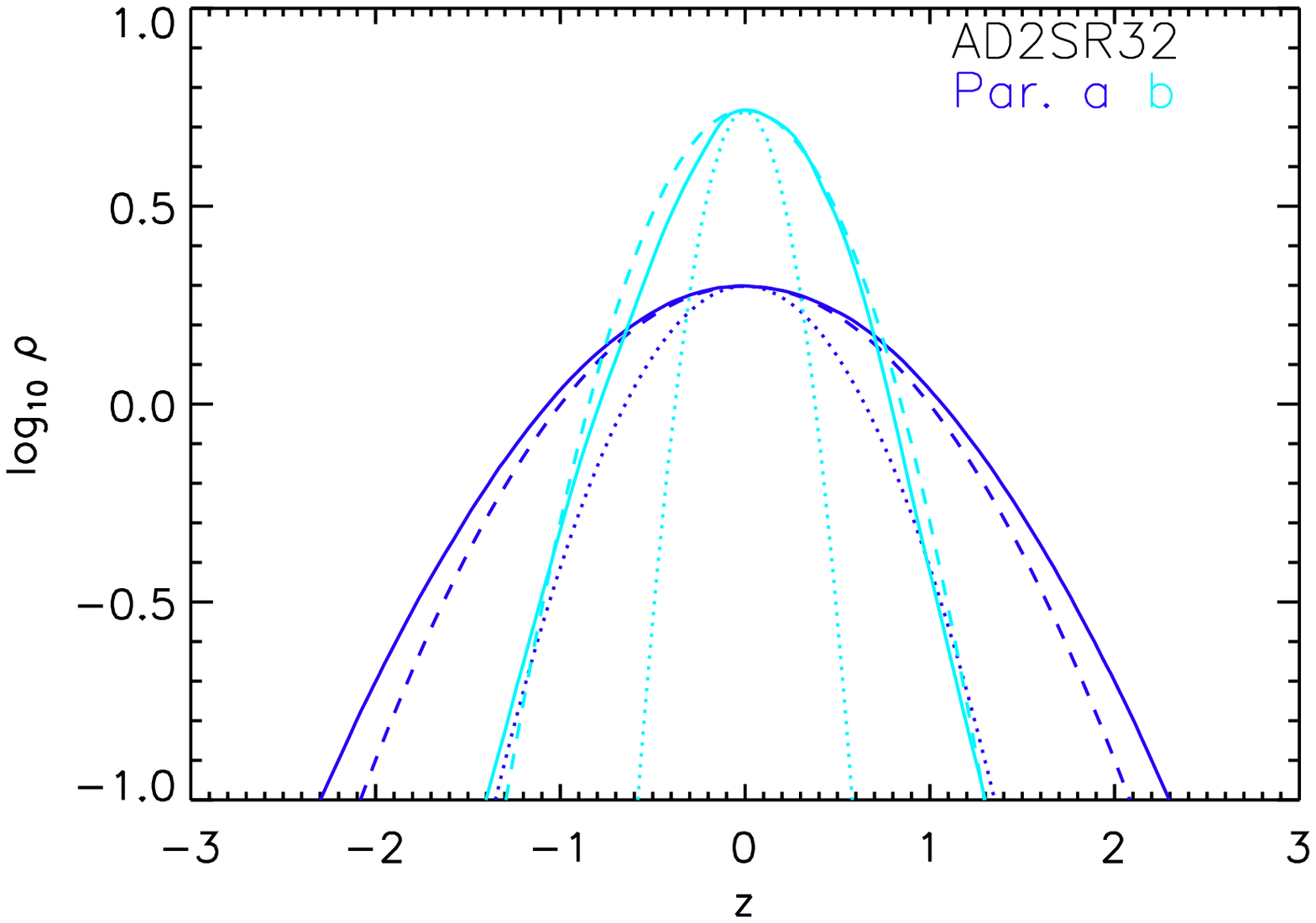} 
\vspace{-0.3 cm}
\caption{Vertical density profiles for Par. a and b in the stratified simulation. The solid curves are from simulations while
the dotted and dashed curves are derived from $h_{d}$ in Equation 
(\ref{eq:hds3}) using the midplane $\langle v_{z}^{2}\rangle$.
The dashed curve assumes $t_{eddy,z}=3 \Omega^{-1}$, while the dotted curve assumes $t_{eddy,z}=(2\Omega)^{-1}$. 
A large $t_{eddy,z}$ fits the density profiles in simulations better. } \label{fig:vertden}
\end{figure}

{ It is possible that
the long-lived structure in $v_{z}$ shown in Figure \ref{fig:spacetime} 
is an artifact of the unstratified shearing box
with periodic boundary conditions in the $z$ direction.} 
To study if our conclusions on particle settling will be changed with stratification in the gas,
we have carried out stratified shearing box simulations.  
The gas disk has an initial Gaussian density profile in the vertical direction. 
The simulation domain for most simulations is [-2h, 2h]$\times$[-2h, 2h]$\times$[-4h, 4h] in the
$x$, $y$, and $z$ directions.
The boundary condition in the $z$ direction is
the same as Simon \etal (2013) which extrapolate both density and magnetic fields from the last active zone to the ghost zones. 
The grid resolution is 32 cells per $h$. The disk is threaded by a net vertical field having $\beta_{0}=10^{4}$. 
For each
type of particles, there are 10$^{6}$ particles uniformly distributed in the box. 
Other simulation parameters and diagnostics are the same as in our unstratified simulations, unless they are specified below.

{ Three simulations in ideal or non-ideal MHD with AD have been carried out}:
(1) The ideal MHD simulation (VSR32);
(2) The AD dominated simulation having Am=1 in the whole box (AD2SR32);
(3) The layered disk simulation (AD2SLR32) with
\begin{equation}
 {\rm Am}={\rm Am}_{0}\times\left(e^{\Sigma_{c}/\Sigma_{+}}+e^{\Sigma_{c}/\Sigma_{-}}\right)
\end{equation}
where $\Sigma_{+}$ and $\Sigma_{-}$ are the integrated surface density above and below 
each grid cell (i.e., $\Sigma_{+}=\int_{z}^{\infty}\rho dz$, $\Sigma_{-}=\int_{-\infty}^{z}\rho dz$),
 $\Sigma_{c}=0.02 \Sigma$, and Am$_{0}$=0.5. The layered disk setup is to
 simulate the minimum-mass solar nebular (MMSN) at 30 AU ($\Sigma=10$ g cm$^{-2}$) with an active layer of $\Sigma_{c}=0.2$ g cm$^{-2}$ ionized by FUV (Bai 2014). 

We divide the disks into the disk midplane within $z=h$ and the disk atmosphere from $h$ to $3h$. 
To measure diffusion coefficients in each region, we trace all particles that stay in that region from 30 to 40 orbits, and use
Equations (\ref{eq:ddz}) and (\ref{eq:ddx}) to derive the diffusion coefficients. 
Mean squared velocities and stresses are measured in the same way as in unstratified simulations.
All these quantities in different regions are given in Table 3. 

{At the disk midplane within $z=h$, $t_{eddy,z}$ is  again larger than one in the
stratified AD simulations ($t_{eddy,z}=2.2$ in AD2SR32, and $t_{eddy,z}=3.5$ in AD2SLR32).
Since the vertical gravity is almost zero at the disk midplane, the turbulence at the disk midplane of stratified disks
should be similar to that in unstratified disks unless there is a feedback loop between the disk midplane and the disk atmosphere. 
Indeed, the measured eddy time at the disk midplane of these stratified simulations are consistent with those in unstratified simulations. }

{Despite $t_{eddy,z}$ at the midplane of AD runs is larger than 1, $t_{eddy,z}$ at  the disk atmosphere 
is  smaller than 1 in these disks.  Considering AD2SR32 and AD2SLR32 have totally different properties in their atmospheres
(Am=1 for AD2SR32 and ideal MHD for AD2SLR32), the similarity in $t_{eddy,z}$ at the atmosphere implies that $t_{eddy,z}$
at the atmosphere is determined by turbulence generated at the midplane of these disks. We suspect that the coherent disturbance coming from the disk
midplane cannot maintain its structure when it is propagating into the low density disk atmosphere, and the turbulence at the disk atmosphere is modulated by the
disk epicyclic motion or vertical oscillation.  }

{ The vertical density profiles for Par. a and b in run AD2SR32 are given in Figure \ref{fig:vertden}.} 
The solid curves are from simulations while
the dotted and dashed curves are derived from Equation (\ref{eq:hds3}) using the midplane $\langle v_{z}^{2}\rangle$.
The dashed curve assumes $t_{eddy,z}=3 \Omega^{-1}$, 
while the dotted curve assumes $t_{eddy,z}=(2\Omega)^{-1}$. 
Clearly the large $t_{eddy,z}$ leads to a better fit to the density profiles in the stratified simulations. 
 
{ To show the large eddy
time in stratified AD simulations (e.g., run AD2SR32),
we plot the space time diagram for the surface density and mean
 velocities along the $x$ direction, as in Figure \ref{fig:stratified}. } The velocities are
averaged over $y$, and $[-h, h]$ in $z$. While
$v_{z}$ does not have a coherent value during 100 orbits in contrast to the unstratified simulations (Figure \ref{fig:unstratified}.), it still has a longer correlation time than $v_{x}$, implying
$t_{eddy,z}>t_{eddy,x}$. Another noticeable feature in Figure \ref{fig:stratified} is the large-scale zonal flows in both density and $v_{y}$ panels, which 
will be discussed in \S 6.2.

\section{Discussion}

\subsection{The Eddy Time and Schmidt Number}
\begin{figure}[ht!]
\centering
\includegraphics[width=0.5\textwidth]{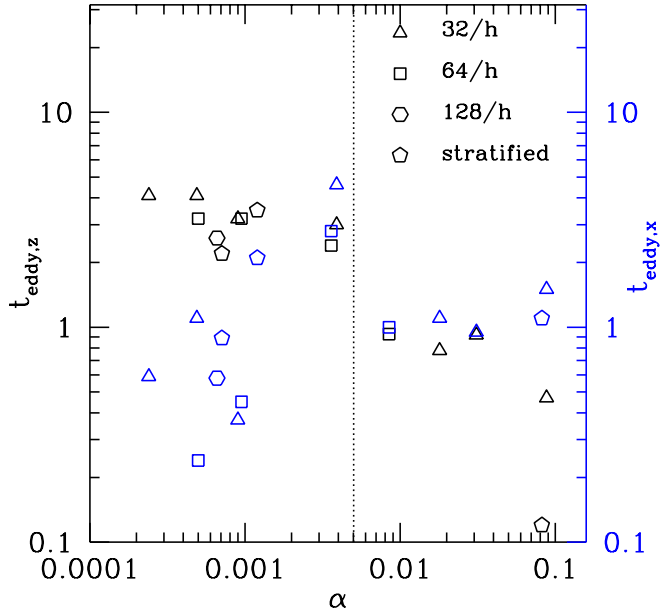} 
\vspace{-0.3 cm}
\caption{ $t_{eddy, x}$ (blue points) and $t_{eddy, z}$ (black points) for all our shearing box simulations. The dotted line
separates ideal MHD simulations (on the right) from MHD simulations with AD (on the left). Clearly, $t_{eddy, z}\sim$3
 for AD runs, while both $t_{eddy,x}$ and $t_{eddy,z}$ $\sim$ 1 for ideal MHD runs.} \label{fig:teddy}
\end{figure}

\begin{figure}[ht!]
\centering
\includegraphics[width=0.5\textwidth]{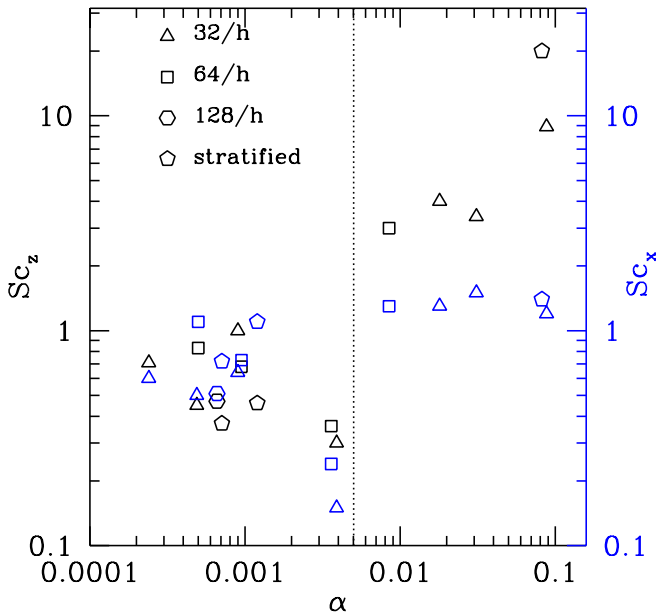} 
\vspace{-0.3 cm}
\caption{ $Sc_{x}$ (blue points) and $Sc_{z}$ (black points) for all our shearing box simulations. The dotted line
separates ideal MHD simulations (on the right) from MHD simulations with AD (on the left).  Clearly, $Sc_{z}$
is smaller than 1  for AD runs, while $Sc_{z}\gtrsim 3$ for ideal MHD runs. } \label{fig:schmit}
\end{figure}

{ The eddy time, also called the integral time, is one of the most important parameters to quantify turbulence,
and it directly relates to turbulent diffusion.} { Figure \ref{fig:teddy} summarizes $t_{eddy}$ in all our simulations. }
For turbulence driven by the MRI with ideal MHD, both $t_{eddy,x}$ and $t_{eddy,z}$ $\sim \Omega^{-1}$. $t_{eddy,z}$ can 
become smaller than $\Omega^{-1}$ only when the disk
is threaded by a strong net vertical magnetic field. For turbulence driven
by the MRI with AD, $t_{eddy,x}$ can range from $5 \Omega^{-1}$ to $0.2 \Omega^{-1}$ as the net magnetic field strength decreases. 
$t_{eddy,z}$ is close to $3 \Omega^{-1}$ in both unstratified and stratified simulations. 

{ Another important parameter to quantify turbulent diffusion is the Schmidt number ($Sc$) which is normally defined as the ratio
between the rates of (angular) momentum transport and mass diffusion. 
We want to caution that there are a few different definitions
of Schmidt number in the literature: Cuzzi \etal (1993) and Youdin \& Lithwick (2007) define Schmidt number as the ratio between the 
gas and dust diffusion coefficient. In fluid dynamics the Schmidt number is defined as the ratio of the viscous (momentum) diffusion
to mass diffusion. While in turbulent disks, the Schmidt number is normally defined as 
the ratio between the R-$\phi$ stress (determining angular momentum
transport) and dust diffusion coefficient (Johansen \& Klahr 2005; Carballido \etal 2005, 2011). In this work we follow the last definition
which has been widely used in numerical simulations to study dust diffusion.
We only show the Schmidt number for gas fluid or well coupled particles. For particles that are less coupled to the gas, the relationship between 
their dust diffusion coefficients and gas diffusion coefficient has been shown in Figure \ref{fig:diffmulti}.}

We calculate $Sc_{x}\equiv \alpha H^{2}\Omega/D_{g, x}$
and $Sc_{z}\equiv \alpha H^{2}\Omega/D_{g,z}$ shown in Table 3 and Figure \ref{fig:schmit}. 

{  In ideal MHD runs,
  $Sc_{x}\sim 1$ and $Sc_{z}\gtrsim 3$,
while in AD runs, both $Sc_{x}$ and $Sc_{z}$ are $\lesssim 1$.} This again suggests that there is 
a qualitatively difference between turbulence in ideal and non-ideal MHD with AD.

\subsection{Particle Trapping In zonal flows}
{ Zonal flows (ZF) are axisymmetric density structures that extend over a large radial range in  MRI turbulent disks.} 
Due to the presence of these density structures, the azimuthal velocity in disks starts to deviate from
a Keplerian rotation profile (e.g.,  $\delta v_{y}$ in Figures \ref{fig:unstratified} and \ref{fig:stratified}) \footnote{Thus the term zonal flows}. 
ZF have been observed in both
local shearing box (Johansen \etal 2009; Simon \etal 2012; Dittrich \etal 2013) 
and global simulations (e.g.,
Dzyurkevich et al. 2010; Flock et al. 2011; Uribe et al.
2011). Although they are more apparent in stratified disk simulations,
they do exist in unstratified simulations as well (e.g., Lyra \etal 2008; Zhu \etal 2013). 

\begin{figure*}[ht!]
\centering
\includegraphics[width=1.\textwidth]{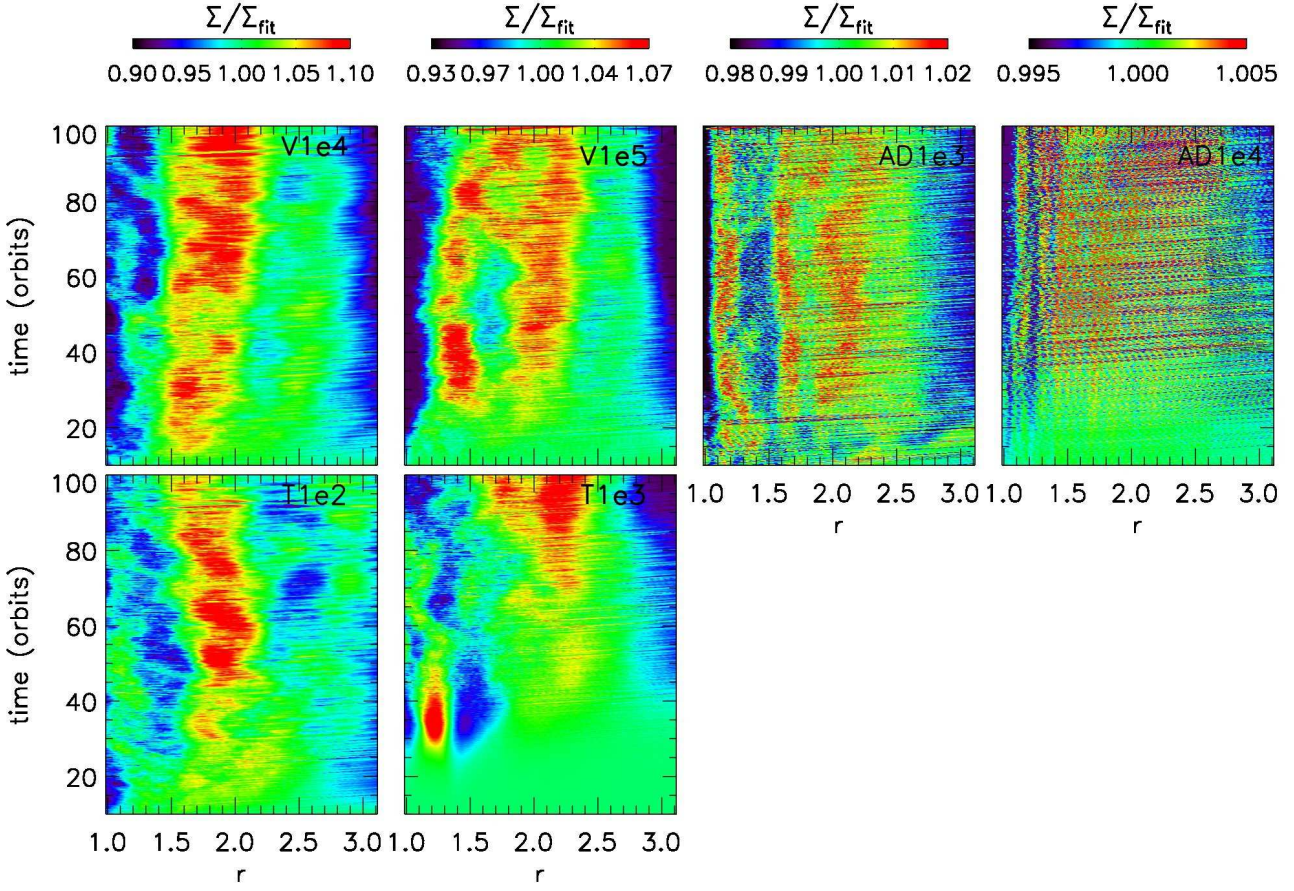}
\caption{The space-time diagram of the normalized gas surface density for various cases. The strength of the disk turbulence decreases in the cases towards right. The amplitude
of the zonal flow also decreases towards right. }\label{fig:multifit}
\end{figure*}

{To show the amplitude of the zonal flows in our simulations, we have plotted the space-time diagrams of the normalized gas surface density
in Figure \ref{fig:multifit}. At each time step, we fit a linear function for ln $\Sigma$ and ln $r$ for disks at $r\in$[1,3] to derive the smooth background density, and then
normalize the gas surface density with this background density. Figure  \ref{fig:multifit} shows that zonal flows in ideal MHD runs have larger amplitude and width than
zonal flows in AD runs. The amplitudes of the zonal flow ($\delta \Sigma/\Sigma_{fit}$) in various cases are: $\sim$ 0.1 for V1e4 and T1e2 (having $\alpha\sim 0.03$),
$\sim$ 0.07 for V1e5 and T1e3 ($\alpha\sim 0.02$),  $\sim$ 0.02 for AD1e3 ($\alpha\sim 2\times 10^{-3}$), and $\sim$ 0.005 for AD2.5e4 ($\alpha\sim 6\times 10^{-4}$).
Thus,the ratio between the ZF amplitude ($\delta \Sigma/\Sigma_{fit}$) and $\alpha$ is  $\sim$3 for ideal MHD runs and $\sim$10 for AD runs. }

{The width of the zonal
flow is 0.5-1 (5-10 disk scale heights) in ideal MHD runs, which is consistent with ideal MHD shearing box  
(Johansen \etal 2009; Simon \etal 2012; Dittrich \etal 2013) and global (Lyra \etal 2008) simulations.  However, in AD runs, the zonal flow is significantly narrower with
 0.2-0.5 (2-5 h) in AD1e3 and 0.1 (1 h) in AD1e4. }

{However, we caution that, based on local shearing-box simulations,  ZF are quite different between  unstratified and stratified disks.} In unstratified disks, 
 both local and global simulations with AD have ZF with small amplitude density fluctuations ( $<5\%$ in Figure \ref{fig:unstratified}). 
On the other hand, when the disk is stratified, 
Figure \ref{fig:stratified} shows much wider and stronger ZF with an amplitude of
$\sim 30\%$ and a radial extent as large as the box size,  consistent with Simon \& Armitage (2014). 

ZF may have a significant impact on planetesimal and planet formation since they can trap dust particles (Dittrich et al. 2013; Simon \& Armitage 2014). As shown in Figure \ref{fig:radialpaper},
 for Par. d, there are significant density concentrations ($\sim$ a factor of $3$ increase of the density) 
 occurring on the scale of $\sim 0.3 r$.

Particle trapping by ZF in global disks is normally considered to be inefficient
when the density variation associated with ZF is insufficient to overcome the background
radial pressure gradient. 
Here we develop a slightly more quantitative
estimate for particle trapping by ZF, and point out that this criterion is not accurate and 
particle trapping by ZF can be quite efficient even in some disks
with low amplitude ZF and a large background radial pressure gradient. 

As discussed in \S 4.1, particle trapping by ZF can be modeled using the 
simple 1-D particle drift diffusion equation (Equation (\ref{eq:sigd})) 
as long as the gas surface density evolution is known from MHD simulations. This serves as a framework for our quantitative estimates 
on particle trapping in ZF as follows.
 
First, we assume that there is a gas density peak due to ZF on top of the smooth background surface density 
$\Sigma_{g,b}(r)$. The disk surface density is 
 $\Sigma_{g}(r)=\Sigma_{g,b}(r)\epsilon(r)$, where
in the peak $\epsilon(r)>1$. Then, assuming the gas radial velocity is much smaller than the dust drift velocity,
we have $v_{g,r}=0$, and
Equation (\ref{eq:driftv}) can be written as
\begin{equation}
v_{d,r}=\frac{v_{K} \frac{c_{s}^2}{r\Omega^2}\frac{\partial {\rm ln} (\Sigma_{g,b} c_{s}^{2})}{\partial r}\left(1+\frac{\partial {\rm ln} \epsilon/ \partial {{\rm ln}r}}{\partial {\rm ln} \Sigma_{g,b} c_{s}^2/ \partial {{\rm ln}r}}\right)}{T_{s}+T_{s}^{-1}}\,.
\end{equation} 
If we denote the drift speed due to the background pressure gradient as $v_{d,r,b}$, we have
\begin{equation}
v_{d,r}=v_{d,r,b}\left(1+\frac{\partial {\rm ln} \epsilon/ \partial {{\rm ln}r}}{\partial {\rm ln} \Sigma_{g,b} c_{s}^2/ \partial {{\rm ln}r}}\right)\,.\label{eq:vdr}
\end{equation} 
Since the radial profiles of the disk's background density and temperature normally follow power laws, $v_{d,r,b}$ can also be 
written as a power law ($v_{d,r,b,0}(r/r_{0})^{\gamma}$).
Combining Equation (\ref{eq:vdr}) and Equation (\ref{eq:sigd}), assuming dust diffusion
is negligible,  and normalizing the equation with drift timescale $-r_{0}/v_{d,r,b,0}$, we find
\begin{equation}
\frac{\partial \Sigma_{d}}{\partial \bar{t}}-\frac{r_{0}}{r}\frac{\partial}{\partial r/r_{0}}\left[\frac{r}{r_{0}}\Sigma_{d}\left(\frac{r}{r_{0}}\right)^{\gamma}\left(1+\frac{\partial {\rm ln} \epsilon/ \partial {{\rm ln}r}}{\partial {\rm ln} \Sigma_{g,b} c_{s}^2/ \partial {{\rm ln}r}}\right)\right]=0\,. \label{eq:sigmad}
\end{equation}
where $ \bar{t}=-t v_{d,r,b,0}/r_{0}$ is the time normalized by the drift timescale. From this equation,
we can see that, if $\gamma$ is a constant,  dust with different sizes follows the same
 surface density evolution with respect to the normalized time $\bar{t}$, independent on 
$v_{d,r,b,0}$ and the particle size. 

Without considering ZF ($\epsilon=1$), Equation (\ref{eq:sigmad}) can be solved analytically 
using the method of characteristics (similar to Youdin \& Shu 2002). With our initial condition of $\Sigma_{d}=\Sigma_{d,0} (r/r_{0})^{-1}$, the solution is
\begin{equation}
 \Sigma_{d}(r, \bar{t})=\Sigma_{d,0}\left(\frac{r_0}{r}\right)^{1+\gamma}\left(\left(\frac{r}{r_{0}}\right)^{1-\gamma}+(1-\gamma)\bar{t}\right)^{\gamma/(1-\gamma)} \,.\label{eq:sigmadr}
\end{equation}

With our assumed disk structure and Equation (\ref{eq:driftv}), 
particles with $T_{s}<1$ have  $\gamma=-1/4$ while particles with
$T_{s}>1$ have $\gamma=1/4$.
Thus, at the drift timescale of $t=-r_{0}/v_{d,r,b,0}$ or $\bar{t}=1$, we have $\Sigma_{d}(r_{0}, 1)=0.85 \Sigma_{d,0}$ for small particles having $\gamma=-1/4$, 
and $\Sigma_{d}(r_{0},1)=1.21 \Sigma_{d,0}$ for big particles having $\gamma=1/4$. 
Thus, the surface density of dust changes less than 20$\%$
within the particle drift timescale in our setup. On long timescales, 
$\Sigma_{d}$ changes at the rate of $t^{-1/5}$ and $t^{1/3}$ respectively, indicating that
 dust accumulation or depletion from the global drift
is inefficient in this case.

Under the condition that
 the background pressure gradient leads to little dust surface density change (i.e., $\gamma<1/2$),  
shallow ZF, which cause $\gamma$ deviate from 0 within ZF, could lead to a significant particle concentration.
With out unstratified disk setup,  the background pressure gradient can lead to fast particle drift
but  won't change the disk surface density significantly ($\gamma$=-1/4 or 1/4 in Equation \ref{eq:sigmadr}). In this case, pressure bumps
in our simulations can lead to significant particle concentration. 
 As shown in Figure \ref{fig:radialpaper}, 
ZF in our ideal MHD simulations can lead to a factor of $\sim$ 2-3 dust density enhancement at density peaks for Par. d during Par. d's drift timescale 
($\sim$ 100 orbits). Based on our dimensionless equation (Equation \ref{eq:sigmad}), 
if the zonal flow can last for $\sim$ 1000 orbits, Par. e could also be concentrated by a factor of $\sim$ 2-3.
Thus, 
particle trapping by small amplitude ZF 
can be efficient, as long as ZF can persist over the particle drift timescale and the background global
pressure gradient leads to little dust surface density change. 

{On the other hand, if $\gamma>1/2$, global particle drift itself can lead to
large dust surface density change. Then the dust surface density change caused by
weak zonal flows is not comparable to the change due to background pressure gradient, and particle trapping by pressure bumps is less significant.}

\subsection{Particle Clustering}

\begin{figure}
\centering
\includegraphics[width=0.5\textwidth]{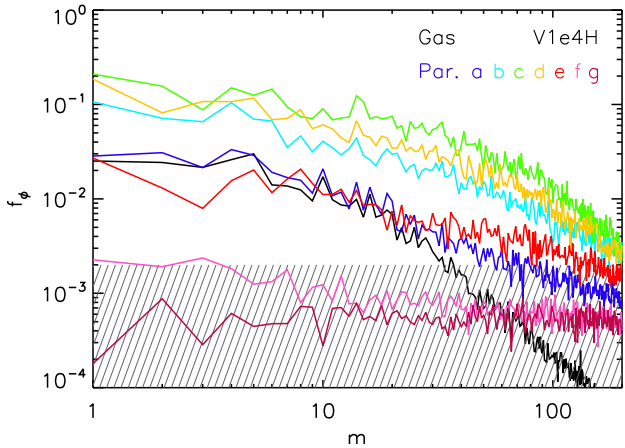} 
\vspace{-0.3 cm}
\caption{Spatial power spectrum of the disk surface density in the azimuthal direction.
The black curve is for the gas surface density, while the colored curves
are for the dust surface density. With increasing particle size from Par. a to
Par. c, the power spectrum increases, indicating dust clumping is more significant when $T_{s}$
is closer to 1. The power spectrum of very large particles (Par. e) can be even lower than
the power spectrum of the gas disk.   The shaded region is dominated by the Poisson noise from
limited number of particles in each grid cell.
} \label{fig:fourierMHD}
\end{figure}

{ Particles drift not only radially due to axisymmetric zonal flows but also azimuthally in response to 
non-axisymmetric gas structures induced by MRI turbulence.} In turbulent disks, the gas density  has non-zero Fourier components
in the $\phi$ direction in  all $m-$modes. In order to study how particles
respond to these azimuthal gas fluctuations, we do Fourier 
transform for the surface density at $r=1$ in the $\phi$ direction 
to get the power spectrum of the dust surface density. 
We want to caution that, due to the particle treatment of the dust, the accuracy of the dust surface density
is limited by the Poisson noise of the number of particles within each grid cell. 
Thus, we use simulation V1e4H which has 0.3 billion particles for each particle type. 
In order to get good statistics at low $m$, we also average the power spectra derived at every orbit from 54 to 60 orbits.
The final power spectrum ($f_{\phi}$) for both gas and dust is shown in Figure \ref{fig:fourierMHD} after 
being normalized by $f_{\phi}(m=0)$.
Even with such a large number of particles, the power spectrum below $f_{\phi}\lesssim 10^{-3}$
 is still dominated by Poisson noise. For example, 
Par. f and g have flat power spectra
which are  typical  for Poisson noise.

{ Figure \ref{fig:fourierMHD} shows that  smaller $m$ modes have higher power, and
particles with different sizes have different power.} Par a.
couples with the gas so well that it almost has the same power spectrum as the gas (black solid curves). With increasing particle size from Par. b 
to Par. c, the power spectrum gets stronger, suggesting stronger particle concentration at all scales. The maximum
power spectrum is achieved for particles with $T_{s}\sim1$ (Par. c).
For bigger particles with $T_{s}>1$ (Par.d and e), the power spectra drop again. For the particles with $T_{s}\gg1$ (e.g., Par. e, Par. f), 
the power spectrum is even weaker than that of the gas, suggesting particles have little response to turbulence,
and the disk is quite axisymetric.  
  
\begin{figure}[ht!]
\centering
\includegraphics[width=0.5\textwidth]{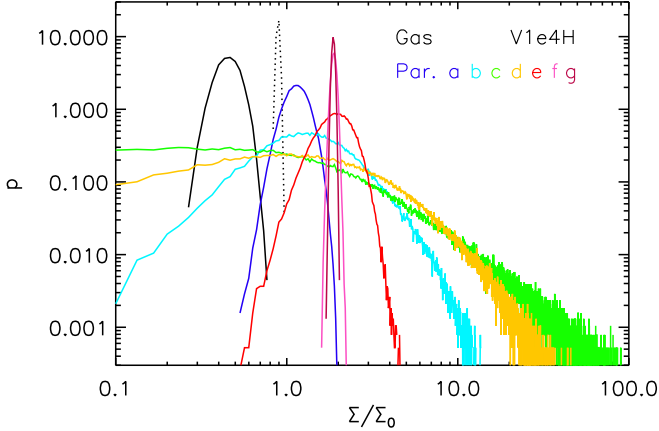} 
\vspace{-0.3 cm}
\caption{The probability distribution function (PDF) for the gas and various types of particles. 
The dotted curve is the PDF for the particle initial condition. The width of the dotted curve
represents the Poisson noise due to the limited number of particles in each grid cell.
} \label{fig:histo}
\end{figure}

{ Another way to study particle concentration at various scales is to calculate the probability distribution function (PDF) for the dust
surface density, as in Dittrich \etal (2013) and Hopkins (2013).} Figure \ref{fig:histo} shows the PDF for both gas and various types of particles
in V1e4H. To compute the PDF, we first divide the surface density of both gas and dust by the initial surface density. 
Then we uniformly
divide the range [0.1,100] into 3000 bins, and among all the grids in the annulus between r=1 and 1.5 we count the fraction of grid cells that have a relative density falling into each bin.
Finally, we divide this value by the bin size, and
 average the resulting functions from each orbit between 54 and 60 orbits to plot Figure \ref{fig:histo}. 

{ Particles with different sizes have different PDFs.}
The dotted curve in Figure \ref{fig:histo}  is the PDF for particles in the initial condition. Thus, the width of this curve
represents the Poisson noise from the limited number of particles within each bin. Any feature in the PDF
comparable to or narrower than this width is unreliable (thus, the PDF of Par. f and g 
is dominated by Poisson noise, similar to Figure \ref{fig:fourierMHD}). The PDF for the gas disk
has a finite width due to turbulence. Par. a couples with the gas very well so that it has a similar PDF
to the gas. Par. b has a much wider PDF, suggesting particles concentrate within turbulent eddies (large $\Sigma$).  
The dust surface density enhancement is most significant for Par. c. For particles with $T_{s}>1$, the PDF narrows again, suggesting the decoupling of dust from gas turbulence. 

 \subsection{Numerical Convergence}
 
\begin{figure*}[ht!]
\centering
\includegraphics[width=0.8\textwidth]{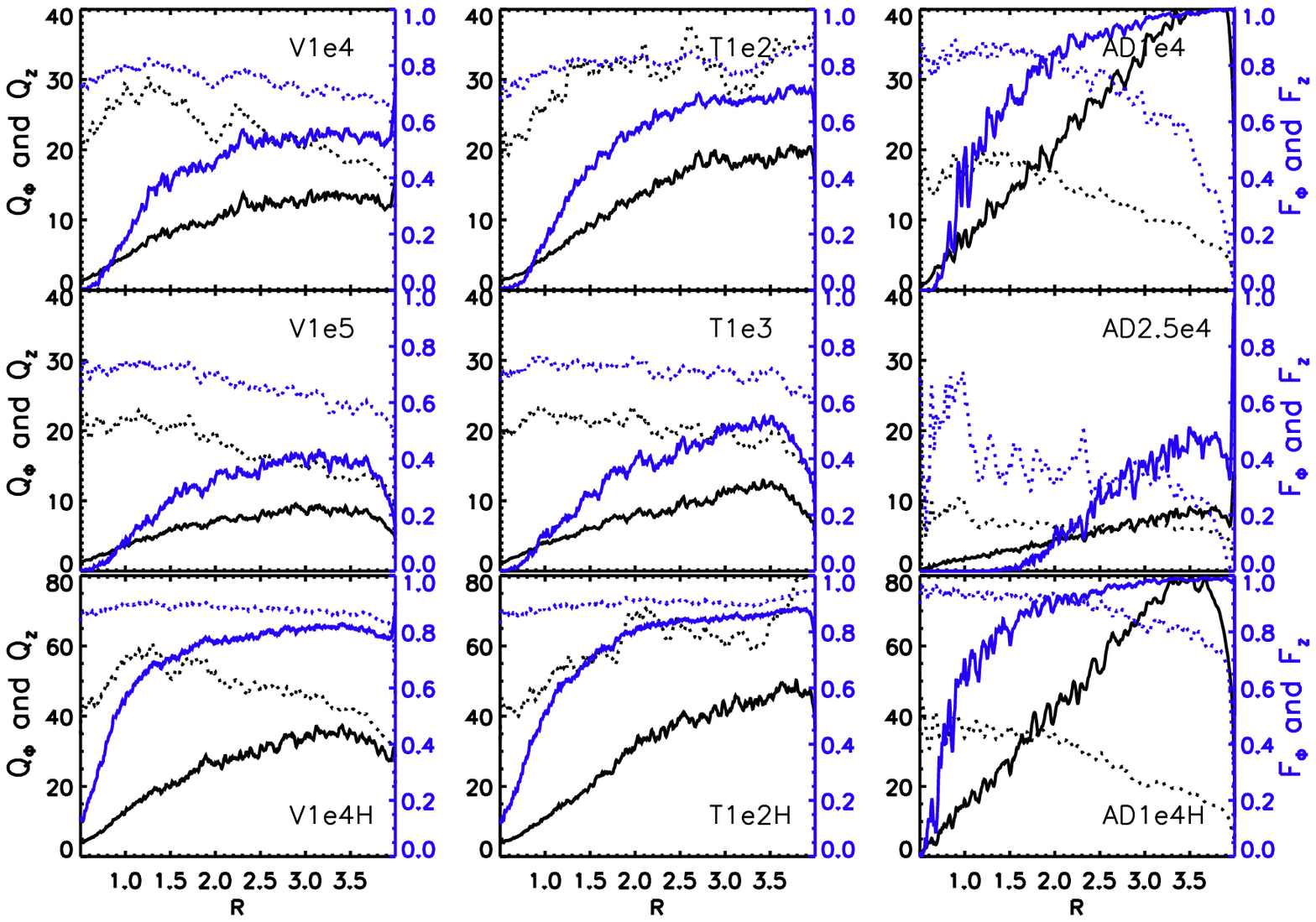} 
\vspace{-0.1 cm}
\caption{The quality factor in the $z$ and $\phi$ directions (Equations (\ref{eq:quality}), $Q_{z}$ (solid black curves), $Q_{\phi}$ (dotted black curves)), 
and the fraction of resolved grids in the $z$ and $\phi$ directions (Equations (\ref{eq:fraction}), $F_{z}$ (solid blue curves),
and $F_{\phi}$ (dotted blue curves)) for various runs at the end of the simulations. } \label{fig:evgqz}
\end{figure*}

To check the numerical convergence of our simulations, we have doubled the numerical resolution from
16/h to 32/h for both ideal and non-ideal MHD with AD. As shown in Table 1, doubling the resolution only
changes the $\alpha$ parameter by less than 10\%, and other parameters--- mean squared velocities, the Reynolds and Maxwell stresses--- are also similar. 

The convergence of the $\alpha$ parameter is only one diagnostic metric among many others 
(quality factors, spectral metrics, tilt angle) to
judge the numerical convergence in global simulations  (Guan \etal 2009; Noble \etal 2010; Beckwith \etal 2011; 
Hawley \etal 2011; Sorathia \etal 2012; Hawley \etal 2013). A thorough study on judging the convergence 
 of global simulations by these metrics is provided in Sorathia \etal (2012) and Hawley \etal (2013).
Sorathia \etal (2012), which has a similar
setup as our unstratified simulations, conclude that
16 grids per scale height is required for the convergence of simulations seeded with a net field. The scale height
defined in Sorathia \etal (2012) is actually $\sqrt{2}$ of the scale height defined in this work. Thus, all our simulations 
fully meet this requirement. 

However, we want to emphasize that, in order to have the correct global accretion structure, 
the simulation needs to be fully converged everywhere in the whole simulation domain. Normally,
some regions in global disks are more resolved than others due to the grid spacing, and
the global structure of the density, temperature and magnetic fields. For example, with uniform grid spacing 
and $h\propto r^{5/4}$, our standard cases have the resolution of 7$/h$ in the $r$ and $z$ directions at the inner boundary
and 90$/h$ at the outer boundary. In order to quantify the resolvability at different radii in disks, we
plot the quality factors $Q$ and $F$ as a function of $r$ at the end of these simulations. 

The quality factor $Q$ is defined as the number of grid cells that resolve the fastest MRI growing mode (Noble \etal 2010),
\begin{eqnarray}
Q_{z}=\lambda_{MRI}/\Delta z\,,\nonumber\\
Q_{\phi}=\lambda_{c}/(r\Delta \phi)\,,\label{eq:quality}
\end{eqnarray}
where $\lambda_{MRI}\equiv 2\pi v_{A,z}/\Omega=8.886 \beta_{z}^{-1/2} h$ in ideal MHD, and  in MHD with AD
\begin{equation}
\lambda_{MRI}\approx 10.26\left(1+\frac{1}{{\rm Am}^{2}}+\frac{1}{{\rm Am}^{1.16\epsilon}}-0.2\epsilon\right)^{1/2}\beta_{z}^{-1/2}
\end{equation}
where $\epsilon\equiv {\rm Am}/(1+{\rm Am})$ (Wardle 1999, Bai \& Stone 2011). With ${\rm Am}=1$,
$\lambda_{MRI}= 17.47 \beta_{z}^{-1/2} h$. $\lambda_{c}$ is defined in the same way as $\lambda_{MRI}$  but using
$\beta_{\phi}$. The quality factor is averaged in both $\phi$ and $z$ directions at each $r$. In Figure \ref{fig:evgqz}, 
the solid black curves are $Q_{z}$, while the dotted black curves are $Q_{\phi}$. Sorathia \etal (2012)
have shown that if $Q_{\phi}\approx 10$, $Q_{z}$ needs to be $\gtrsim 10-15$, and if $Q_{\phi}\gtrsim 25$,
$Q_{z}$ can be smaller ($\sim$ 6 in their Figure 8). Figure \ref{fig:evgqz} shows that even our standard simulations
 have $Q_{\phi}\sim 20$ and $Q_{z}\gtrsim 5$
in most regions, except those close to the inner boundary where vertical magnetic fields 
are lost through the inner boundary. 

Another diagnostic metric proposed by Sorathia \etal (2012) is the fraction $F$ of
grids that resolve the fastest growing modes by at least eight grid cells. 
\begin{eqnarray}
F_{z}(r)=\frac{\int (\lambda_{MRI}\ge 8\Delta z) rd\phi dz}{\int rd\phi dz}\,,\nonumber\\
F_{\phi}(r)=\frac{\int (\lambda_{c}\ge 8R\Delta \phi) rd\phi dz}{\int rd\phi dz}\,,\label{eq:fraction}
\end{eqnarray}
where the logical statement within the integral takes the value of one 
if the statement is true, and zero if the statement is false. The $F$ factors from our simulations 
are shown in Figure \ref{fig:evgqz} as the blue curves. Solid blue curves are $F_{z}$, and
the dotted blue curves are $F_{\phi}$. For most of our standard runs (except AD2.5e4), we have $F_{z}\gtrsim 0.4$
and $F_{\phi}\gtrsim 0.6$, which means that around half of the grids have resolved the MRI
fastest growing modes and the simulations should be converged (Sorathia \etal 2012).
For our high resolution runs, $F_{z}$ and $F_{\phi}$ are larger than 0.8 in most regions,
indicating most regions in our simulations have been fully resolved.

\section{Conclusion}
We have studied dust transport in turbulent protoplanetary disks using
three-dimensional magnetohydrodynamic (MHD) simulations 
including Lagrangian dust particles. 
The turbulence is driven by the magnetorotational instability (MRI) 
in either ideal  or non-ideal MHD with ambipolar
diffusion (AD). Our aim is to test if the evolution and vertical structure of the dust disk in global 3-D
MHD simulations can be reproduced by simple 1-D models. If simple 1-D models
can be justified, the evolution of dust can be studied in long timescales without the need of expensive
3-D MHD simulations. 

\begin{itemize}
\item In ideal MHD runs, we confirm that the dust radial diffusion coefficient measured in simulations agree well with the analytical formulae
in Youdin \& Lithwick (2007) with the assumption of $t_{eddy,x}\sim t_{eddy,z}\sim \Omega^{-1}$.

\item Both the surface density evolution and vertical structure of the dust disk in 3-D global unstratified simulations can be roughly reproduced
by simple azimuthally averaged 1-D and analytical models  (except for dust with $T_{s}\sim 1$).

\item However, in order to capture particle trapping by pressure bumps due to MRI turbulence, we need a more refined
1-D model that uses the evolution of gas surface density from 3-D MHD simulations.

\item There is a noticeable discrepancy between MHD simulations and azimuthally averaged 1-D models  for  surface density of  dust having $T_{s}\sim$1, 
indicating that non-axisymmetric density features in MRI turbulent disks can affect these particles significantly.  

\item In AD runs, turbulence is significantly suppressed, 
and the evolution of dust in our simulations is dominated by particle radial drift. 
The vertical structure of the dust disk can again be fitted by the simple analytical model but it requires a much larger $t_{eddy,z}$
than the value in ideal MHD runs. 

\item By carrying out both unstratified and stratified local
shearing box simulations with Lagrangian particles, we find  that $t_{eddy,z}$ is $\sim 3 \Omega^{-1}$ in our AD runs.
On the other hand, $t_{eddy,x}$ can range from much smaller than to much larger than $\Omega^{-1}$, depending on the
strength of the net magnetic field.  

\item In unstratified AD simulations, we observe that,  at some parts of the disk, $v_{z}$ can be even correlated over the whole
simulation time. This may imply that the linear growing mode still persists in the nonlinear phase and affects particle dynamics.

\item In deal MHD runs, $Sc_{r}\sim 1$ and $Sc_{z}\gtrsim 3$, implying angular momentum transport is more efficient than dust diffusion.
However, in AD runs, both $Sc_{r}$ and $Sc_{z}$ are $\lesssim 1$ implying dust diffusion is more efficient than angular momentum transport
in the outer part of the protoplanetary disk where AD dominates.

\item The difference on particle diffusion between AD and ideal MHD runs is due to that the presence of AD changes 
both the temporal autocorrelation function and power spectrum of turbulence. In AD runs, the temporal autocorrelation function has
a long correlation time for $v_{z}$, while the correlation time for $v_{x}$ can be either long or 
short sensitively depending on the net magnetic field strength. The temporal power spectrum also flattens or even increases
at  high frequencies (or short timescales). 
Since the detailed form of the power
spectrum determines dust diffusion coefficients for particles with $T_{s}\gtrsim 1$,
the formulae given by Youdin \& Lithwick (2007) are not applicable for describing diffusion of
big particles in AD disks. 

\item For global unstratified simulations, the amplitude of zonal flows becomes smaller when the turbulence becomes weaker.
The weak ZF in AD runs cannot trap particles significantly. 
The ratio between the ZF amplitude ($\delta \Sigma/\Sigma$) and $\alpha$ is  $\sim$3 for ideal MHD runs and $\sim$10 for AD runs. 
The width of the zonal
flow is 0.5-1 (5-10 disk scale heights) in ideal MHD runs, while significantly narrower in AD runs.
In our ideal MHD simulations, zonal flows can increase the dust surface density 
by a factor of three as long as zonal flows can
persist over the particle drift timescale.

\end{itemize}
 This study suggests that turbulent disks in non-ideal MHD with AD have dramatically different dust diffusion
 coefficients than those in ideal MHD.
 Future studies on particle diffusion in MRI turbulent disks dominated by 
non-ideal MHD effects is necessary for 
understanding dust transport in realistic protoplanetary disks.

\acknowledgments
We thank the referee for a very helpful report. All simulations were carried out using
computers supported by the Princeton Institute of Computational Science and Engineering and Kraken at National 
Institute for Computational Sciences through XSEDE grant TG-AST130002. 
Z.Z. acknowledges support by
NASA through Hubble Fellowship grant HST-HF-51333.01-A
awarded by the Space Telescope Science Institute, which is
operated by the Association of Universities for Research in Astronomy, Inc., for NASA, under contract NAS 5-26555.

\end{document}